\documentclass[10pt,preprint]{aastex}
\usepackage{color}

\newcommand{\av}{$A_V$}

\newcommand{\etal}{et~al.}

\newcommand{\ks}{$K_{\rm s}$}

\newcommand{\mum}{$\mu$m}

\begin{document}

\title{New Young Star Candidates in CG4 and Sa101}

\slugcomment{Version from \today}

\author{L.\ M.\ Rebull\altaffilmark{1}, 
C.\ H.\ Johnson\altaffilmark{2},
V.\ Hoette\altaffilmark{3}, 
J.\ S.\ Kim\altaffilmark{4},
S.\ Laine\altaffilmark{1},
M.\ Foster\altaffilmark{4},
R.\ Laher\altaffilmark{1},
M.\ Legassie\altaffilmark{1,5},
C.\ R.\ Mallory\altaffilmark{6},
K.\ McCarron\altaffilmark{7}
W.\ H.\ Sherry\altaffilmark{8}
}

\altaffiltext{1}{Spitzer Science Center/Caltech, M/S 220-6, 1200
E.\ California Blvd., Pasadena, CA  91125
(luisa.rebull@jpl.nasa.gov)}
\altaffiltext{2}{Breck School, Minneapolis, MN}
\altaffiltext{3}{Yerkes Observatory, University of Chicago}
\altaffiltext{4}{University of Arizona, Tucson, AZ}
\altaffiltext{5}{Raytheon, Pasadena, CA}
\altaffiltext{6}{Pierce College, Woodland Hills, CA}
\altaffiltext{7}{Oak Park and River Forest High School, Oak Park, IL}
\altaffiltext{8}{NOAO/NSO, Tucson, AZ}

\begin{abstract}

The CG4 and Sa101 regions together cover a region of $\sim$0.5 square
degree in the vicinity of a ``cometary globule'' that is part of the
Gum Nebula.  There are seven previously identified young stars in this
region; we have searched for new young stars using mid- and
far-infrared data (3.6 to 70 microns) from the Spitzer Space
Telescope, combined with ground-based optical data and  near-infrared
data from the Two-Micron All-Sky Survey (2MASS). We find infrared
excesses in all 6 of the previously identified young stars in our
maps, and we identify 16 more candidate young stars based on apparent
infrared excesses.  Most (73\%) of the new young stars are Class II
objects. There is a tighter grouping of young stars and young star
candidates in the Sa101 region, in contrast to the CG4 region, where
there are fewer young stars and young star candidates, and they are
more dispersed.  Few likely young objects are found in the ``fingers''
of the dust being disturbed by the ionization front from the heart of
the Gum Nebula.

\end{abstract}


\section{Introduction}
\label{sec:intro}

Hawarden and Brand (1976) identified ``several elongated, comet-like
objects'' in the Gum Nebula. These objects have dense, dark, dusty
heads and long, faint tails, which are generally pointing away from
the center of the Vela OB2 association. More such ``Cometary
Globules'' (CGs) were subsequently identified in the Gum Nebula (e.g.,
Sandqvist 1976, Reipurth 1983), but similar structures had also been
identified elsewhere (e.g., Orion, Rosette Nebula, etc.) in the context
of Bok Globules (e.g., Bok \& Reilly 1947) and ``elephant trunks''
(e.g., Osterbrock 1957).  These objects are all thought to be related
in the following sense -- certain regions of the molecular cloud are
dense enough to persist when the stellar winds and ionizing radiation
from OB stars powering an \ion{H}{2} region move over them, initially
forming elephant trunks and then eventually cometary globules. These
structures often also have bright rims, thought to originate from the
OB stars' winds and radiation, and are often actively forming stars
(e.g., Reipurth 1983), most likely triggered by the interaction with
winds and radiation from the OB stars (e.g., Haikala \etal\ 2010).

Cometary Globule 4 (CG4) in the Gum Nebula has a striking appearance
(see Figures~\ref{fig:where}--\ref{fig:irac4mosaic}), where the
combination of the original dust distribution plus ablation from the
OB winds and radiation has resulted in a relatively complicated
structure.  A serendipitous placement of background galaxy ESO
257-~G~019 just 0.15$\arcdeg$ to the East of the heart of CG4 adds to
the drama of the image (see
Figures~\ref{fig:where}--\ref{fig:irac4mosaic}).  About a half a
degree to the West of the heart of CG4 is another cloud, named Sa101.
This region was initially recognized by Sandqvist (1977) as an opacity
class 5 (on a scale of 6, e.g., fairly dark) dark cloud. This cloud
appears to have been shadowed, at least partially, by CG4 from the
ionization front. (See discussion in Pettersson 2008.)

Reipurth \& Pettersson (1993) studied the region of CG4+Sa101, finding
several H$\alpha$ emission stars; see Table~\ref{tab:knownysos}.  They
point out that CG-H$\alpha$1 and 7 are not associated with dusty
material, and, as such, may have been associated with clumps that have
already evaporated, as opposed to the stars still projected onto
molecular cloud material.  They argue on the basis of H$\alpha$
equivalent widths that these cannot be foreground or background dMe
stars, but are instead likely young stars, members of the
association.  They then use this to argue that this cloud is most
likely a part of the Gum Nebula Complex (as opposed to a foreground or
background object). We therefore assume as well that the stars
associated with CG4+Sa101 are also associated with the Gum Nebula.

Distances to CG4+Sa101 and even the Gum Nebula are uncertain, with
values between 300 and 500 pc appearing in the literature (e.g., 
Franco 1990).  The generally accepted source of strong ultraviolet
(UV) radiation is $\gamma^2$Velorum, which is taken to be 360-490 pc
away (Pozzo \etal\ 2000). However, the Gum Nebula is elongated along
our line of sight, so the distance to different parts of the nebula
could be significantly different than the distance to
$\gamma^2$Velorum.  Vela OB2 is $\sim$ 425 pc (Pozzo \etal\ 2000).  In
the context of this paper, we test the extrema of the distance
estimates of 300 and 500 pc, though we note that our results are not
strongly dependent on distance. 

Since the CG4+Sa101 region contains some previously identified young
stars, it is likely that there are more young stars, perhaps lower
mass or more embedded than those discovered previously.  Kim \etal\
(2003), using a preliminary reduction of some of the optical data used
here, discussed some additional candidate young stars in this region.
Since it is now commonly believed that every low-mass star goes
through a period of having a circumstellar disk, young stars can be
identified via an infrared (IR) excess, assumed to be due to a
circumstellar disk. A survey in the IR can be used to identify objects
having an IR excess and thus distinguish candidate young stars from
most foreground or background objects, at least those foreground or
background stars without circumstellar disks.   The IR also more
easily penetrates the dusty environs of star-forming regions,
particularly globules  such as these cometary globules in the Gum
Nebula.

The Spitzer Space Telescope (Werner \etal\ 2004) observed the
CG4+Sa101 region with the Infrared Array Camera (IRAC; Fazio \etal\
2004) at 3.6, 4.5, 5.8, and 8 \mum, and with the Multiband Imaging
Photometer for Spitzer (MIPS; Rieke \etal\ 2004) at 24 and 70 \mum. We
used these data to search this region for additional young stellar
object (YSO) candidates. We combined these Spitzer data with data from
the near-infrared Two-Micron All-Sky Survey (2MASS; Skrutskie \etal\
2006) and from ground-based optical photometric data that we have
obtained, and used the multi-wavelength catalog to evaluate and rank 
our list of Spitzer-selected YSO candidates.

The observations and data reduction are described in 
\S\ref{sec:obs}.  We select YSO candidates using Spitzer colors in
\S\ref{sec:findthem}, and discuss their overall properties in
\S\ref{sec:properties}.  We include a few words on the serendipitously
observed galaxy in \S\ref{sec:galaxy}. Finally, we summarize our main
points in \S\ref{sec:concl}.

\begin{deluxetable}{lcclllllllll}
\tablecaption{Previously identified young stars in the
CG4+Sa101 region\tablenotemark{a}\label{tab:knownysos}}
\rotate
\tabletypesize{\tiny}
\tablewidth{0pt}
\tablehead{
\colhead{name}  & \colhead{Region} &  \colhead{RA (J2000)} & \colhead{Dec (J2000)} 
& \colhead{$U$ (mag)}& \colhead{$B$ (mag)}  & \colhead{$V$ (mag)} &
\colhead{$J$ (mag)} &  \colhead{$H$ (mag)} &\colhead{$K$ (mag)} &
\colhead{Spec.~Type} }
\startdata
CG-H$\alpha$ 1\tablenotemark{b} &  SA 101&07 30 37.6 &-47 25 06 &\nodata& \nodata& $>$17  &\nodata&\nodata&\nodata& M3-4 \\
CG-H$\alpha$ 2 &  SA 101&07 30 57.5 &-46 56 11 &\nodata& \nodata& $>$17  &\nodata&\nodata&\nodata& M2: \\
CG-H$\alpha$ 3 &  SA 101&07 31 10.8 &-47 00 32 &17.50  & 16.59  & 14.99  &11.51&10.35&9.62& K7 \\
CG-H$\alpha$ 4 &  SA 101&07 31 21.8 &-46 57 45 &16.91  & 15.99  & 14.59  &11.21&10.38&9.91& K7-M0 \\
CG-H$\alpha$ 5 &  SA 101&07 31 36.6 &-47 00 13 &16.74  & 16.51  & 15.25  &11.73&10.64&9.96& K2-5 \\
CG-H$\alpha$ 6 &  SA 101&07 31 37.4 &-47 00 21 &16.53  & 15.63  & 14.21  &10.42& 9.52&9.06& K7 \\
CG-H$\alpha$ 7 &  CG 4  &07 33 26.8 &-46 48 42 &16.00  & 15.16  & 13.97  &\nodata&\nodata&\nodata& K5 \\
\enddata
\tablenotetext{a}{Information tabulated here comes from Reipurth \&
Pettersson (1993), with positions updated to be J2000 and tied to the
Spitzer and 2MASS coordinate system.  We assumed the errors on the
photometry to be $\sim$20\% when plotting them in the spectral energy
distributions (SEDs) in Figures~\ref{fig:seds1}--\ref{fig:seds3}.}
\tablenotetext{b}{Off the edge of the Spitzer maps discussed here.}
\end{deluxetable}

\section{Observations, Data Reduction, and Ancillary Data}
\label{sec:obs}

\begin{figure*}[tbp]
\epsscale{1}
\plotone{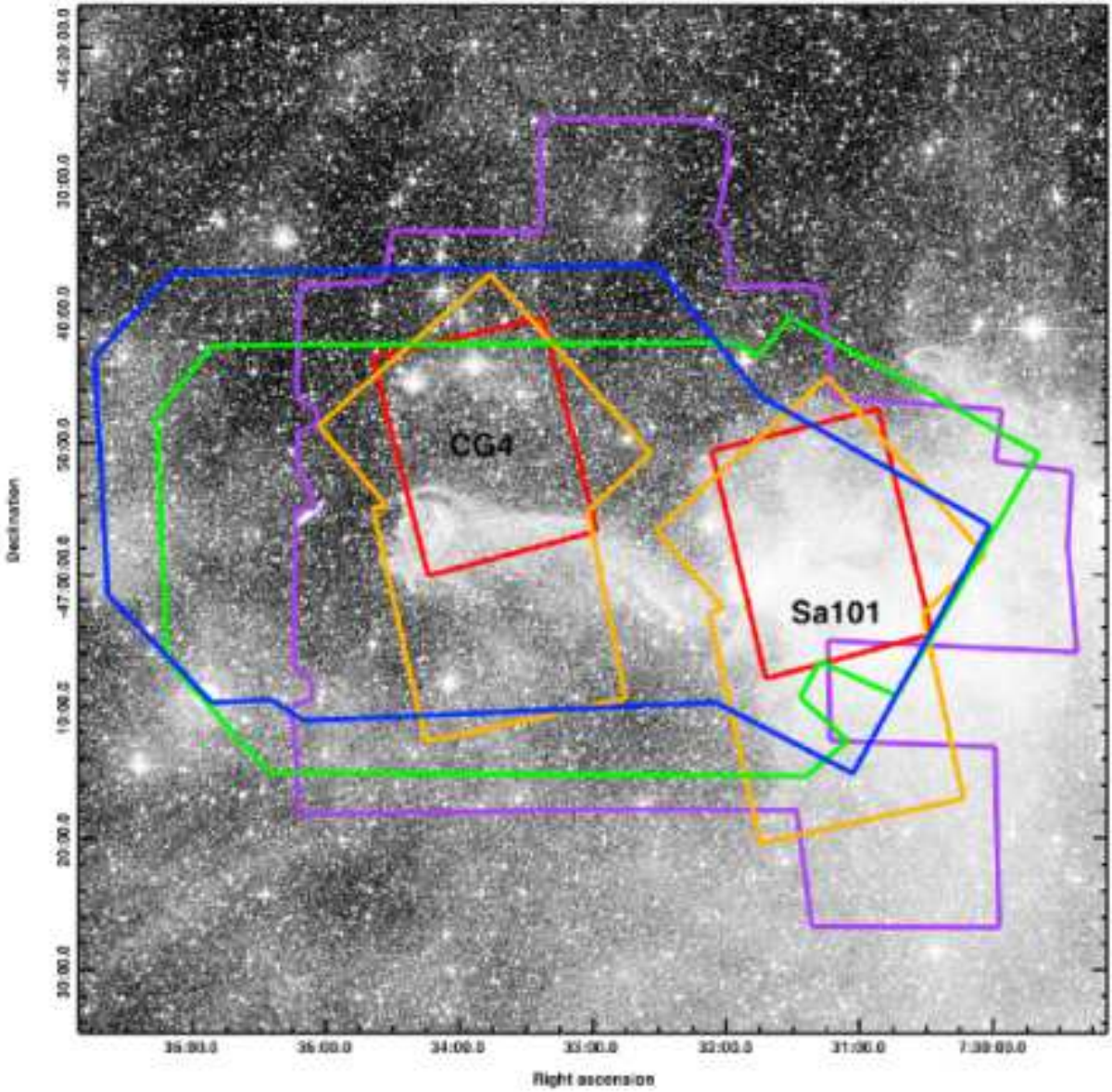}
\caption{Approximate location of optical, IRAC, MIPS coverage,
superimposed on a Palomar Observatory Sky Survey (POSS) image. Purple
is optical, blue is IRAC-1 and -3, green is IRAC-2 and -4, orange is
MIPS-1, and red is MIPS-2. The approximate locations of CG4 and Sa101
are also indicated. The galaxy ESO 257-~G~019 is located on the left,
partially obscured by the edge of the optical coverage.}
\label{fig:where}
\end{figure*}

In this section, we discuss the IRAC and MIPS data acquisition and
reduction.  We briefly summarize the optical ($BVR_cI_c$) data
reduction, which will be covered in more detail in Kim \etal, in
preparation.  We also discuss merging the photometric data across
bands, and with the 2MASS near-IR catalog ($JHK_s$). The regions of the
sky covered by IRAC, MIPS, and the optical observations are indicated
in Figure~\ref{fig:where}.

We note for completeness that the four channels of IRAC are 3.6, 4.5,
5.8, and 8 microns, and that the three channels of MIPS are 24, 70,
and 160 microns. These bands can be referred to equivalently by their
channel number or wavelength; the bracket notation, e.g., [24],
denotes the measurement in magnitudes rather than flux density units
(e.g., Jy). Further discussion of the bandpasses can be found in, e.g., the
Instrument Handbooks, available from the Spitzer Science Center (SSC)
or the Infrared Science Archive (IRSA) Spitzer Heritage Archive (SHA) 
websites.

\subsection{IRAC Data}

\begin{figure*}[tbp]
\epsscale{0.75}
\plotone{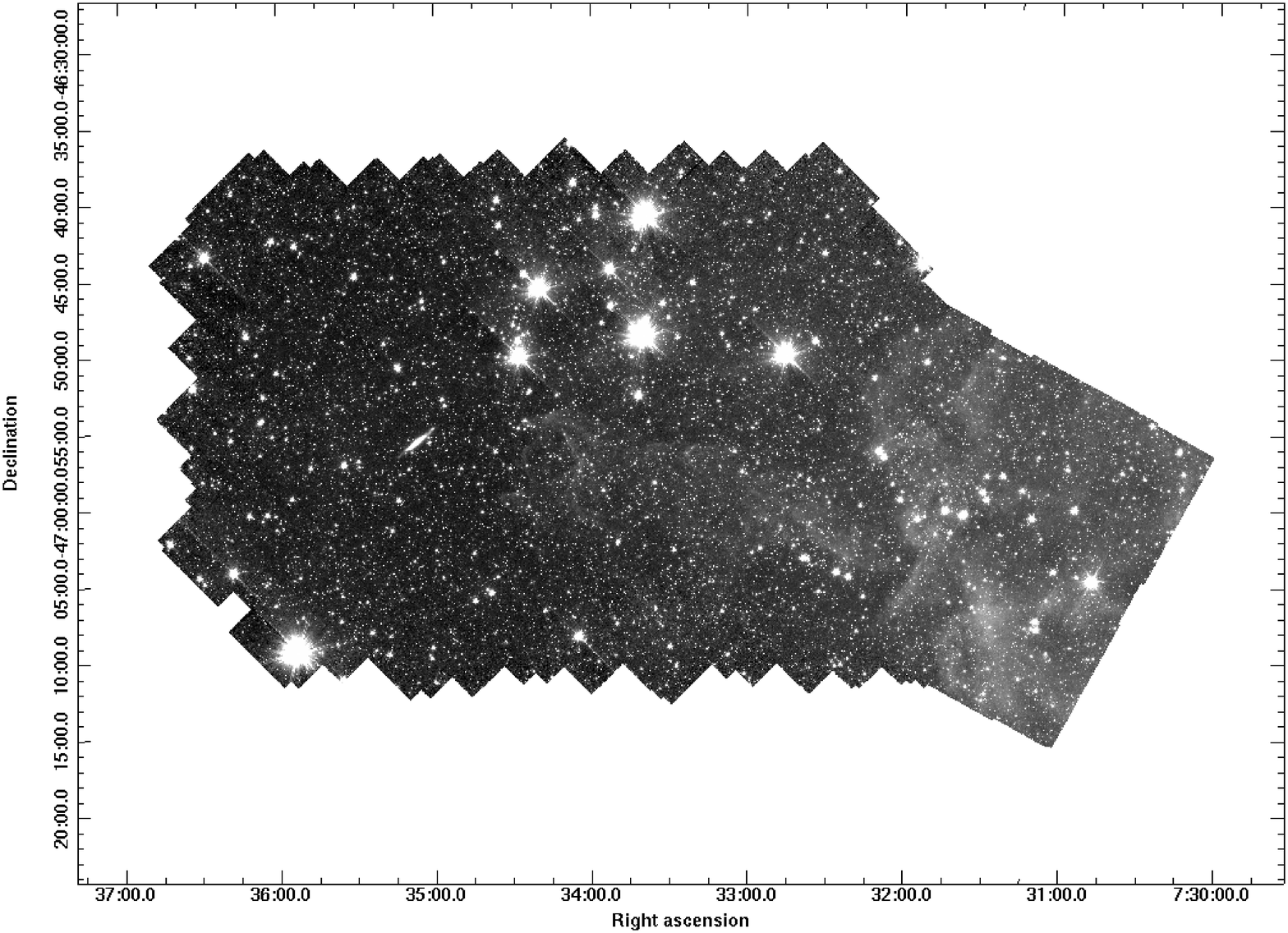}
\caption{The IRAC 3.6 \mum\ (channel 1) mosaic. }
\label{fig:irac1mosaic}
\end{figure*}
\begin{figure*}[tbp]
\epsscale{0.75}
\plotone{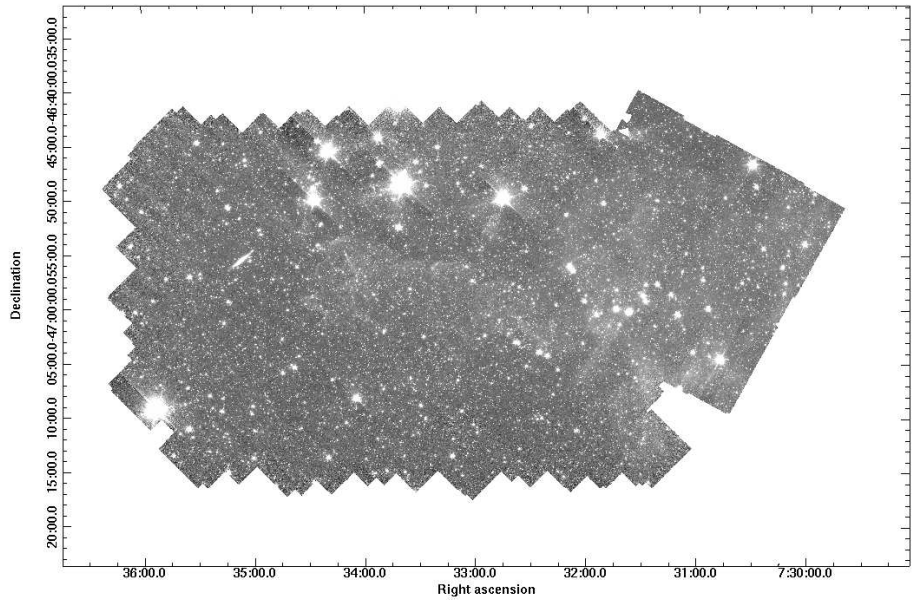}
\caption{The IRAC 4.5 \mum\ mosaic (channel 2). }
\label{fig:irac2mosaic}
\end{figure*}
\begin{figure*}[tbp]
\epsscale{0.75}
\plotone{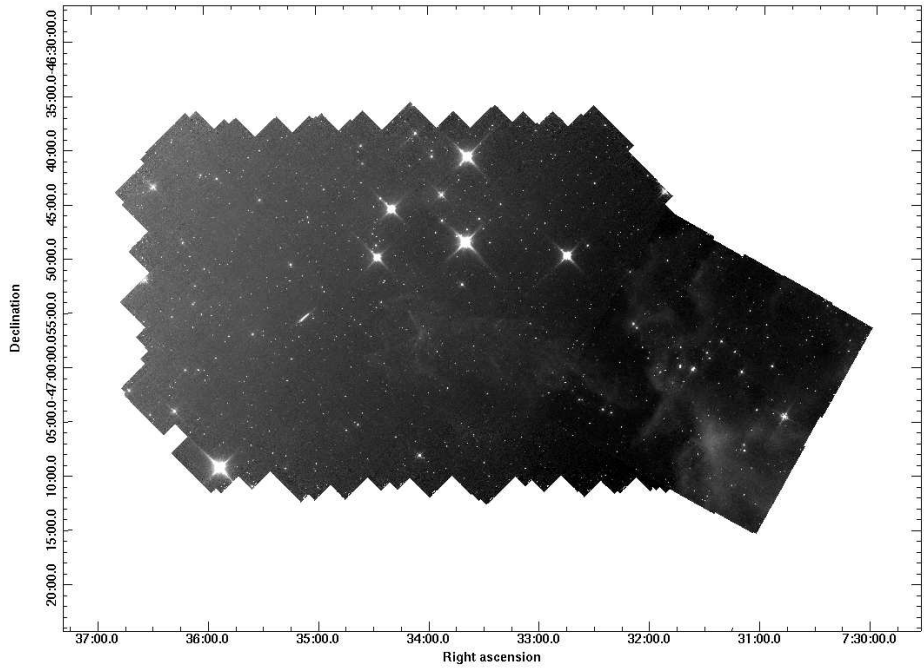}
\caption{The IRAC 5.8 \mum\ mosaic (channel 3). }
\label{fig:irac3mosaic}
\end{figure*}
\begin{figure*}[tbp]
\epsscale{0.75}
\plotone{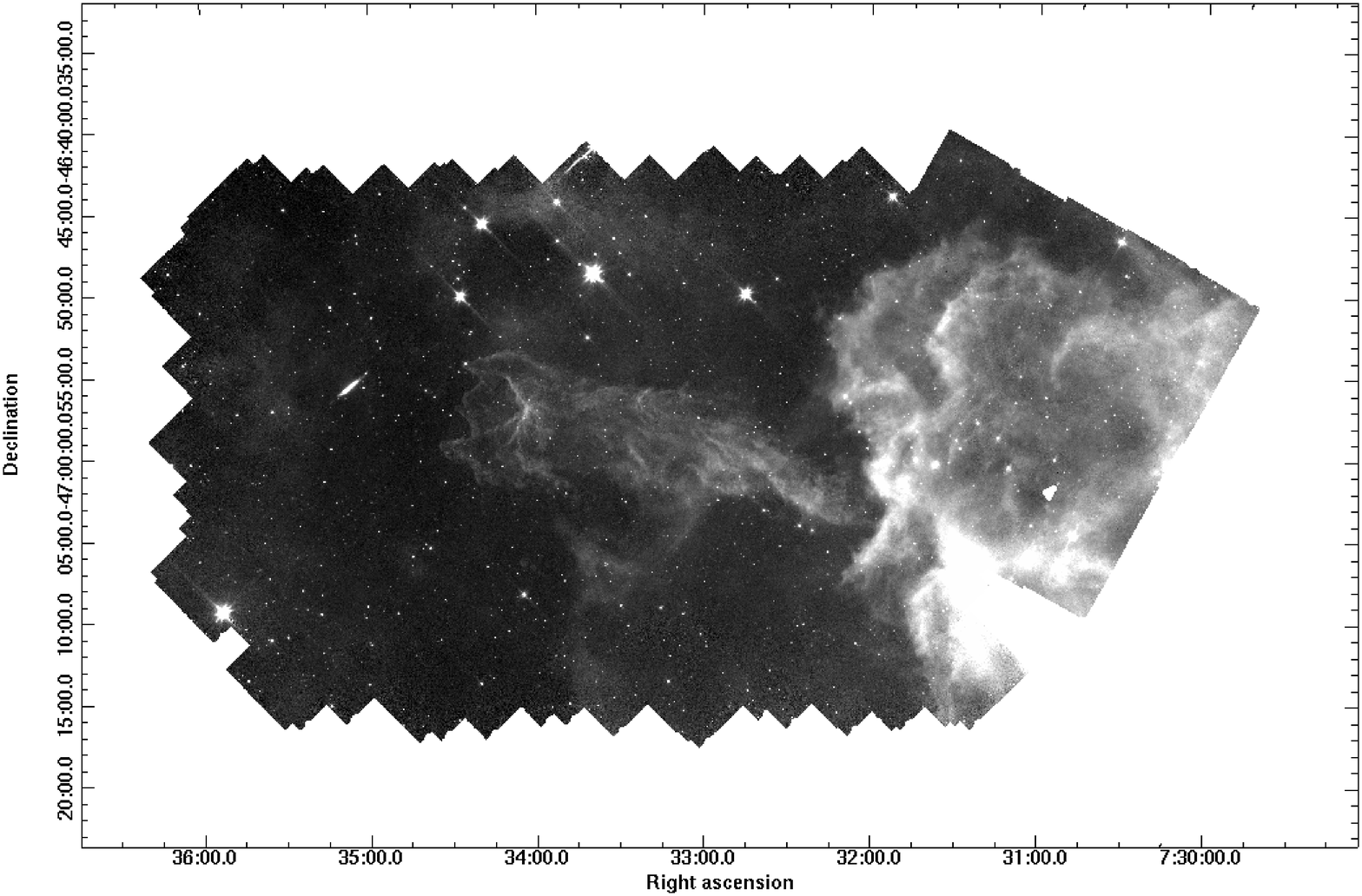}
\caption{The IRAC 8 \mum\ mosaic (channel 4). }
\label{fig:irac4mosaic}
\end{figure*}

We used the IRAC data for CG4 from program 462, AORKEY\footnote{An AOR
is an Astronomical Observation Request, the fundamental unit of
Spitzer observing. An AORKEY is the unique 8-digit identifier for the
AOR, which can be used to retrieve these data from the Spitzer
Archive.} 24250880; for Sa101, we used the IRAC data from program
20714, AORKEY 16805888.  The IRAC data from program 462 (for CG4) were
taken on 2007-12-29 with 12 sec high-dynamic-range (HDR) frames, so
there are two exposures at each pointing, 0.6 and 12 sec, with 3
dithers per position, for a total integration time of 36 seconds (on
average). The IRAC data from program 20714 (for Sa101) were taken on
2006-03-27 with 30 sec HDR frames; there are also two exposures per
pointing, but deeper, 1.2 and 30 seconds.  For this observation, there
are two dithers per position, for a total integration time of 60
seconds (on average).  Because of the different integration times, we
reduced the Sa101 and CG4 observations independently even though they
overlap on the sky (in
Figures~\ref{fig:irac1mosaic}-\ref{fig:irac4mosaic}, the jagged-edged
observation is from the observation in program 462, and the
smooth-edged observation on the right is from the observation in
program 20714).

We note that there are additional IRAC data in this region that we did
not use.  IRAC data from program 202 were of a very small region
centered on the head of the globule. We did not include these data in
an effort to make our survey as uniform as possible over the entire
surveyed region. IRAC data from program 20714 for CG4 were taken in
non-HDR mode, with 30 second exposures; as a result of some very
bright stars in the field of view, the instrumental effects rendered
these data very difficult to work with. Our science goals near CG4 can
be met with the total integration time from program 462 alone.  In
addition, the data from program 462 cover a larger area
($\sim0.8\arcdeg\times\sim1\arcdeg$) than from program 20714 
($\sim0.4\arcdeg\times\sim0.5\arcdeg$).  For these reasons, we did not
incorporate the CG4 IRAC data from program 20714 in this analysis.

We started with the corrected basic calibrated data (CBCDs) processed
using SSC pipeline version 18.7.  Because of the very bright stars in
the field of view, and because the data from program 462 were taken
with cluster targets, we could not use the pipeline-processed
mosaics.  Moreover, the artifact correction, which is normally done
for individual cluster targets separately in the SSC pipeline
processing, is much improved when using the CBCD files from program
462 all at once.  We reprocessed the IRAC data from both program 462
(for CG4) and program 20714 (for Sa101), using MOPEX (Makovoz \&
Marleau 2005) to calculate overlap corrections and create mosaics with
very much reduced instrumental artifacts compared to the pipeline
mosaics.  The pixel size for our mosaics was the same as the pipeline
mosaics, 0.6 arcseconds, half of the native pixel scale. We created
separate mosaics for the long and the short exposures at each channel
for photometric analysis.  For display purposes, we further used MOPEX
to combine the two long-frame observations into one large mosaic per
channel, as seen in
Figures~\ref{fig:irac1mosaic}-\ref{fig:irac4mosaic}.  The component
mosaics were properly weighted in terms of signal-to-noise and
exposure time. The total area covered by at least one IRAC channel (as
seen in Figures~\ref{fig:irac1mosaic}-\ref{fig:irac4mosaic}) is
$\sim$0.5 square degrees. 

To obtain photometry of sources in this region, we used the
APEX-1frame module from MOPEX to perform source detection on the
resultant long and short mosaics for each observation separately.  We
took those source lists and used the aper.pro routine in IDL to
perform aperture photometry on the mosaics with an aperture of 3
native pixels (6 resampled pixels), and an annulus of 3-7 native
pixels (6-14 resampled pixels). The corresponding aperture corrections
are, for the four IRAC channels,  1.124, 1.127, 1.143, \& 1.234,
respectively, as listed in the IRAC Instrument Handbook.  As a check
on the photometry, the educators and students associated with this
project (see Acknowledgments) used the Aperture Photometry Tool (APT;
Laher \etal\ 2011a,b) to confirm by hand the measurements for all the
targets of interest.  To convert the flux densities to magnitudes, we
used the zero points as provided in the IRAC Instrument Handbook:
280.9, 179.7, 115.0, and 64.13 Jy, respectively, for the four
channels. (No array-dependent color corrections nor regular color
corrections were applied.) We took the errors as produced by IDL to be
the best possible internal error estimates; to compare to flux
densities from other sources, we took a flat error estimate of 5\%
added in quadature.

To obtain one source list per channel per observation, we then merged
the short and the long exposures for each channel separately, and for
each observation independently because of the different exposure times
as noted above. The crossover point between taking fluxes from the
short and long exposures were taken from empirical studies of prior
star-forming regions, and were magnitudes of 9.5, 9.0, 8.0, \&  7.0
for the four IRAC channels repsectively. 
We performed this merging via a strict by-position search, looking for
the closest match within 1$\arcsec$.  This maximum radius for
matching was determined via experience with other star-forming regions
(e.g., Rebull \etal\ 2010).  
The limiting magnitudes of these final source lists are the same for
both observations, and are [3.6]$\sim$17 mag, [4.5]$\sim$17 mag,
[5.8]$\sim$15.5 mag, and [8]$\sim$14.5 mag.


\subsection{MIPS Data}

\begin{figure*}[tbp]
\epsscale{0.75}
\plotone{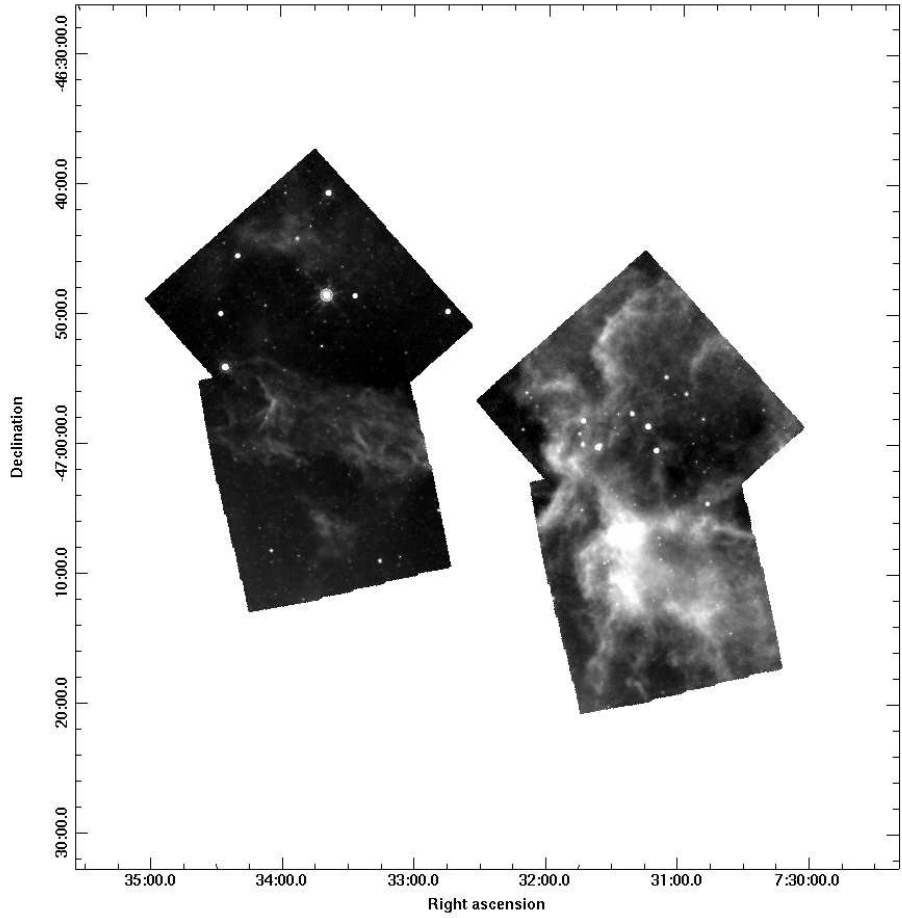}
\caption{The MIPS 24 \mum\ mosaic. The 24 \mum\ coverage consists of
two pointed photometry-mode small maps, plus the 24 \mum\ data
serendipitously obtained during 70 \mum\ photometry observations.   
Extended emission and point
sources are both apparent.}
\label{fig:mips1mosaic}
\end{figure*}
\begin{figure*}[tbp]
\epsscale{0.75}
\plotone{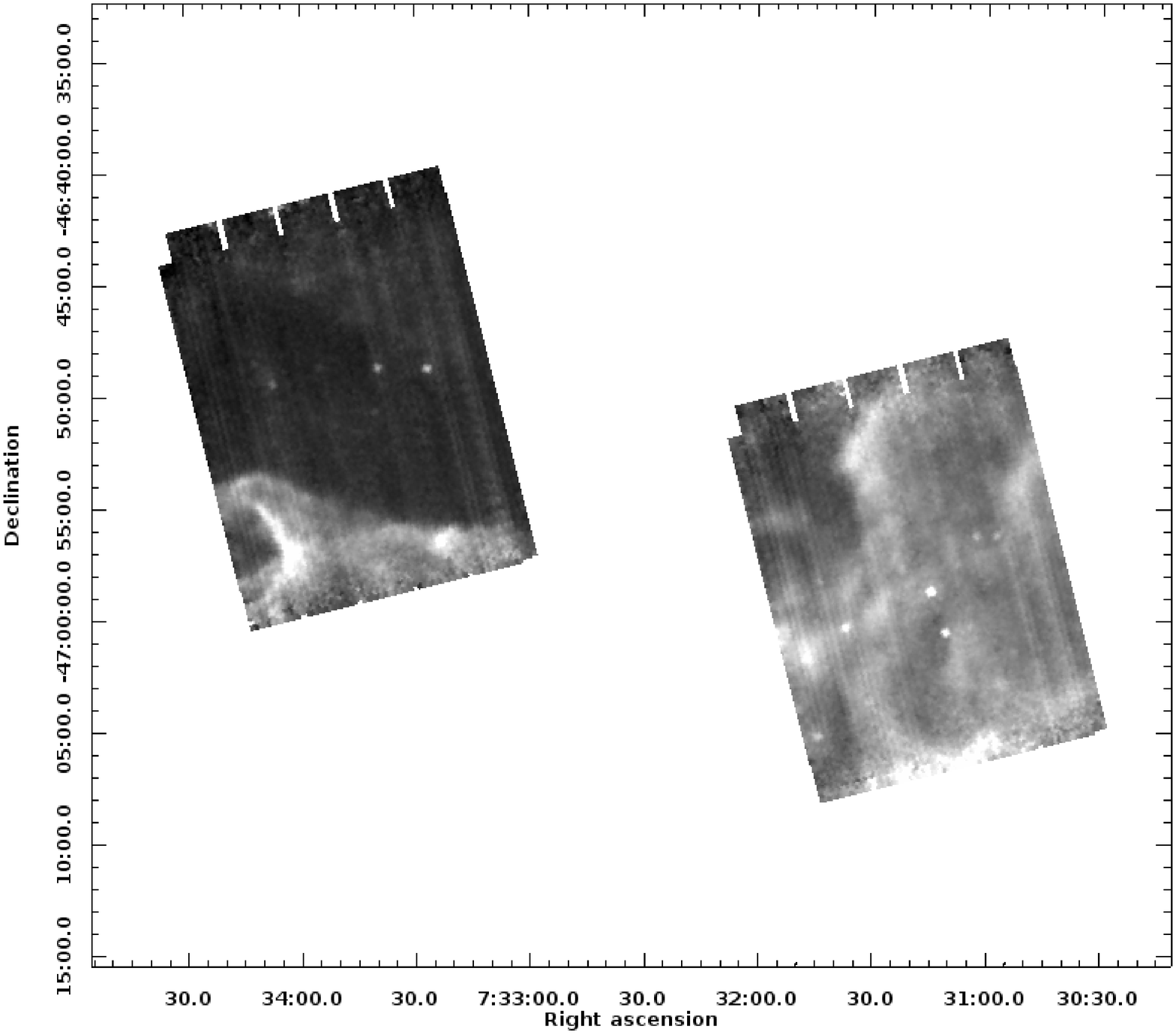}
\caption{The MIPS 70 \mum\ mosaic.  The 70 \mum\ coverage is
two pointed photometry-mode small maps.  Extended emission and point
sources are both apparent. }
\label{fig:mips2mosaic}
\end{figure*}

There are two MIPS AORs in the CG4 region and two MIPS AORs in the
Sa101 region, all four of which were obtained as part of program 20714
(AORKEYs 16805632, 16807936, 16806144, 16808192) on 2006-05-08 or
2006-06-12; see Figures~\ref{fig:where}, \ref{fig:mips1mosaic}, and
\ref{fig:mips2mosaic}.  The AORs were obtained in MIPS photometry
mode, nominally centered on 7:31:18.7, -46:57:45 for Sa101 and 7:33:48
-46:49:59.9 for CG4. One of each pair of AORs is explicitly a MIPS-24
photometry observation, and the other is a MIPS-70 photometry
observation.  During the 70 \mum\ observation, the 24 \mum\ array is
still turned on and is still collecting valid data. We combined the
prime 24 \mum\ data from the MIPS-24 photometry mode observations with
the serendipitous 24 \mum\ data from the MIPS-70 photometry mode
observations to obtain larger maps at 24 \mum. The original explicitly
24 \mum\ photometry observations are small photometry-mode maps, with
3 s integration per pointing, but a net integration time of $\sim$210
s over most of the resultant mosaic. The serendipitously obtained data
averaged $\sim$350 s net integration time; where the deliberate and
serendipitous data overlapped, integration times can be $\sim$450 s. 
The 70 \mum\ photometry observations were also small photometry-mode
maps, with 10 s integration per pointing; the net integration time
over most of the mosaics was $\sim$400 s.

The data for 24 \mum, like those for IRAC, were affected by the bright
objects, and required additional processing beyond what the online
pipelines could provide. We started with S18.13 enhanced BCDs (eBCDs)
from the pipeline. We implemented a self-flat for each AOR separately,
as described in the MIPS Instrument Handbook, available from the SSC
or IRSA SHA 
websites.  For each pair of overlapping 24 micron maps, we then ran an
overlap correction using the overlap script that comes with MOPEX, and
then created one mosaic for CG4 and one for Sa101, again using MOPEX.
Our mosaics had the same pixel size as the online mosaics,
2.45$\arcsec$.   In order to combine the images into one mosaic for
display in Figure~\ref{fig:mips1mosaic}, the different overall
background levels between the two observations (having an origin in
the different Zodiacal light levels at the times of the two
observations) were problematic. The brighter of the two was
artificially lowered via median subtraction to bring its dynamic range
into a similar regime as the fainter; this renders photometry on this
net mosaic invalid, but the morphology seen in
Figure~\ref{fig:mips1mosaic} is still valid.  The total area covered by
the net 24 \mum\ map is only $\sim$0.3 square degrees, smaller than
that of the IRAC map. 

To obtain photometry at 24 \mum, we ran APEX-1frame on each of the
mosaics (one per observation) and performed point-response-function
(PRF) fitting photometry using the SSC-provided PRF.  Tests using the
apex\_qa module portion of MOPEX suggest that our photometry is well
within expected errors.  For three problematic sources, we used
aperture photometry instead of the PRF-fitted photometry, as they
provided a better fit in apex\_qa.  We used the signal-to-noise ratio
(SNR) value returned by APEX-1frame as the best estimate of the
internal (statistical) errors, adding a 4\% flux density error in
quadrature as a best estimate of the absolute uncertainty.  The
limiting magnitude of these observations is [24]$\sim$10.5 mag.  Note
that we optimized our data reduction to obtain measurements of the
brighter sources and sources superimposed on the nebulosity; many
sources fainter than this are apparent in the image but not included
in our catalog, simply because our scientific goals are aimed at the
brighter objects.  For one source of interest below (073243.5-464941,
which was considered and then rejected as a YSO candidate; see
\S\ref{sec:findthem}), an upper limit was obtained at the given
position by laying down an aperture as if a source were there, and
taking 3 times that value for the 3$\sigma$ limit.   To convert the
flux densities to magnitudes, we used the zero point as found in the
MIPS Instrument Handbook, 7.14 Jy.


At 70 \mum, there are viable observations from AORKEYs 16807936 and
16808192.  We downloaded data processed with pipeline version S18.12. 
The online pipeline does a very good job of producing mosaics; see
Figure~\ref{fig:mips2mosaic}, where there are a handful of point
sources and extended emission visible. The online pipeline produces
both filtered and unfiltered mosaics; the filtering preserves the flux
densities of the point sources and improves their signal-to-noise,
especially for faint sources, but destroys the flux density
information for the extended emission. The unfiltered mosaics are
shown in Figure~\ref{fig:mips2mosaic}, but we performed photometry on
the filtered mosaics. The pipeline mosaics have resampled 4$\arcsec$
pixels (as opposed to 5.3$\arcsec$ native pixels), and the two
observations together cover about 0.1 square degrees.   We used
APEX-1frame to do PRF fitting on the pipeline filtered mosaics for the
point sources, using the SSC-provided PRF. For one problematic source,
aperture photometry provided a better flux density estimate.  There
are only 11 objects with 70 \mum\ detections, and there is a large
variation in background levels, so quoting a limiting magnitude is
difficult, but is very approximately 3 mag.
We assumed a conservative, flat 20\% flux density error.  The zero
point we used again came from the MIPS Instrument Handbook, 0.775 Jy.

Where there was 70 \mum\ coverage for the sources of interest, 
we placed an aperture at the expected location of the source, and performed 
photometry as if there were a source there, taking 3 times that value as
the 3$\sigma$ limit that appears in Table~\ref{tab:ourysos} below.

\subsection{Optical Data}

The optical data will be discussed further in Kim \etal, in
preparation, but we summarize the important aspects of the data
reduction here.

The $BVR_cI_c$ photometry of the CG4+Sa101 region were obtained during
2001 March 6, 7, 9, 10, and 11 using the 2K$\times$2K CCD at the 0.9m
telescope at the Cerro Tololo Inter-American Observatory (CTIO).  The
images have a pixel scale of 0.4\arcsec in a 13.6\arcmin field of view.

Bias and twilight sky flat fields were taken at the beginning and
at the end of each night. Long and short (300 sec and 30 sec)
exposures were taken for object fields. During every photometric
night we observed Landolt (1992) standard stars of two or three fields
several times per night for photometric calibration.

We performed aperture photometry using multiple aperture sizes, and
the photometry with highest signal-to-noise (S/N) ratio were chosen as final
photometry for each star. For the standard stars, we used an aperture
radius of 17 pixels. We used IRAF\footnote{IRAF is distributed by the
National Optical Astronomy Observatory, which is operated by the
Association of Universities for Research in Astronomy (AURA) under
cooperative agreement with the National Science Foundation.}/PHOTCAL
routines to solve for the zero point, extinction and color terms of
the standard star solution.

For the target stars in CG4 and Sa101, we used a custom IDL photometry
pipeline (written by W.~H.~Sherry), which was developed for the CTIO
0.9m telescope. For each target, aperture photometry was performed
using multiple size apertures starting from aperture size of 2 pixels
to 17 pixels. The highest S/N photometry was chosen as the final
photometry.

We used an aperture correction to place our photometry on the same
photometry system as our standard stars. The point-spread function
(PSF) of the CTIO 0.9m telescope varies noticeably as a function of
location on the CCD. This is insignificant for large apertures, but
for aperture sizes of 2-3 pix, the difference can be a few percent. We
accounted for the spatial dependence of the aperture correction in
each image by fitting the aperture corrections for stars with
photometric errors better than 0.02 mag with a quadratic function. The
uncertainty of the aperture correction is about 0.01 magnitude. For
each aperture, the uncertainty on the instrumental magnitude is
calculated including the uncertainty on the aperture correction for
each aperture.

We used {\it imwcs} program written by D.~Mink\footnote{Documentation
and source code are available at
http://tdc-www.harvard.edu/software/wcstools/} (Mink 1997) to
determine the astrometric solution of each image. It fits the pixel
coordinates of our targets to the known positions of USNO A2.0 stars
located in each field.  We then measured positions to an accuracy of
$\sim$0\farcs3 relative to the USNO A2.0 reference frame.

For each pointing, we used a 1$\arcsec$ matching radius to match
sources within the optical $BVR_cI_C$ filters. The coordinates are
averaged over different filters, and the typical average uncertainty
is 0\farcs2--0\farcs3.  To find duplicates between adjacent
pointings, we again looked for positional matches within 1$\arcsec$,
and then took a weighted mean of the available photometry. 

Basically the entire IRAC map was covered west of 113.8 degrees RA
(07:35:12); see Figure~\ref{fig:where}. We have found no YSO
candidates east of this. The completeness limits over the field are as
follows: $B\sim19$ mag, $V\sim18$ mag, $R_c\sim17.5$ mag, $I_c\sim17$
mag.  However, we note that there are fewer sources per projected area
detected (e.g., effectively shallower limits) in the regions where
there is molecular cloud material. The zero-points we used for
conversion of the magnitudes to flux densities (for inclusion in the
spectral energy distributions in
Figures~\ref{fig:seds1}--\ref{fig:seds3}) were, respectively, 4000.87,
3597.28, 2746.63, and 2432.84 Jy.

\subsection{Bandmerging}
\label{sec:bandmerging}

In summary, to bandmerge the available data, we first merged the
photometry from all four IRAC channels together with near-IR 2MASS 
data within each observation (CG4 and Sa101), and then merged together
those source lists from each observation.  We next included the MIPS
data, and then the optical data.  We now discuss each of these steps
in more detail.  At the end of this section, we discuss some aggregate
statistics of the bandmerged catalog.

To merge the photometry from all four IRAC channels together, we
started with a source list from 2MASS.  This source list includes
$JHK_s$ photometry and limits, with high-quality astrometry.   We merged
this source list by position to the IRAC-1 source list, using a search
radius of 1$\arcsec$, a value  empirically determined via experience
with other star-forming regions (e.g., Rebull \etal\ 2010).  Objects
appearing in the IRAC-1 list but not the $JHK_s$ list were retained as
new potential sources.  The master catalog was then merged, in
succession, to IRAC-2, 3, and 4, again using a matching radius of
1$\arcsec$. 

Because we are primarily interested in objects detected by Spitzer, we
dropped any objects not having flux densities in at least one Spitzer
band (e.g., objects off the edge of the Spitzer maps, having
measurements only in 2MASS).  Because the source detection algorithm
we used can be fooled by instrumental artifacts, we also explicitly
dropped objects seen only in one IRAC band as likely artifacts.

For the sources of interest later in the paper, most have counterparts
in all three bands in 2MASS by this point in the merging. However, for
two sources, this matching failed, at least in part. For one source,
073121.8-465745 (=CG-Ha4), the automatic merging found a counterpart
with a measured $J$ magnitude, but the $HK_s$ measurements are flagged
with a photometric quality (ph\_qual) flag of `E', denoting that the
goodness of fit quality was very poor, or that the photometry fit did
not converge, or that there were insufficient individual data frames
for the measurement. Since there was a good measurement at $J$, we
assumed that there were sufficient frames at $HK_s$, and that
something had happened to the fit.  We took the values as reported in
the catalog, assumed a large error bar, and used these values in the
table and plots below; as will be seen in the spectral energy
distribution (SED) for this object (Fig.~\ref{fig:seds2}, top center),
these values are probably close to what is most appropriate for this
object. For source 073425.3-465409, the automatic merging fails, most
likely because the 2MASS counterpart is slightly extended, and it may
be slightly extended at 3.6 \mum\ as well. The nearest match in the
catalog is $\sim2\arcsec$ away, very large by comparison to other
source matches here, but manual inspection strongly suggests that this
is the appropriate counterpart to the source seen at Spitzer bands.
See Appendix~\ref{app:073425.3-465409} for more on this interesting source.


We then compared the source lists from the separate observations, CG4 and
Sa101, again using a matching radius of 1$\arcsec$. For objects
detected in more than one mosaic, we took a weighted average of the
flux density at the corresponding band.

Next, we merged the 70 \mum\ source list to the 24 \mum\ source list.
The 70 \mum\ point-spread function is large compared to the positional
accuracy needed, and astrophysically, each 70 \mum\ source ought to
have a counterpart at 24 \mum, given the sensitivity of these
observations.  We individually verified that each of the 70 \mum\
point sources had a 24 \mum\ counterpart, and then merged these two
source lists. To successfully have the computer match the sources that
were clearly matches by eye, a positional accuracy of 2.5$\arcsec$ was
required, consistent with our experience in other star-forming
regions (e.g., Rebull \etal\ 2010).   Since the two MIPS observations
do not overlap with each other, no explicit merging of the MIPS source
lists from the two observations was required beyond simple
concatenation.

To combine the merged MIPS source list into the merged 2MASS+IRAC
catalog, we used a positional source match radius of 2$\arcsec$, again
determined via experience with other star-forming regions (e.g.,
Rebull \etal\ 2010).  The MOPEX source detection algorithm can be
fooled by structure in the nebulosity in the image, and by inspection,
this was the case for these data. To weed out these false `sources',
we then dropped objects from the catalog that were detections only at
24 \mum\ and no other bands.  Finally, to merge the $J$ through 70
\mum\ catalog to the optical ($BVR_cI_c$) catalog, we looked for nearest
neighbors within 1$\arcsec$. 

After this entire process, there are $\sim$21,000 sources with IRAC-1
(3.6 \mum) or IRAC-2 (4.5 \mum) detections, $\sim$9000 sources with
IRAC-3 (5.8 \mum) detections, and $\sim$4000 sources with IRAC-4 (8
\mum) detections. About 3000 ($\sim$15\%) of the IRAC sources have
viable data at all 4 IRAC bands, nearly all of which have counterparts
in 2MASS.  The optical data do not cover the entire IRAC map, but
about half of the 4-band IRAC detections have counterparts in the
optical catalog. There are only $\sim$500 sources at 24 \mum\ in our
catalog and just 11 sources at 70 \mum; note that the MIPS-24 map
covers a much smaller area than the IRAC maps, and the MIPS-70 map is
smaller still (see Figures~\ref{fig:where}-\ref{fig:mips2mosaic}). Ten
of the 11 MIPS-70 sources have counterparts at all four IRAC bands;
the one that does not is saturated at 2 of the IRAC bands. About 200
sources have all four IRAC bands plus MIPS-24.


\section{Selection of YSO candidates with infrared excess}
\label{sec:findthem}

With our new multi-wavelength view of the CG4+Sa101 region, we can
begin to look for young stars. We focus on finding sources having an
infrared excess characteristic of YSOs surrounded by a dusty disk.
There is no single Spitzer color selection criterion (or set of
criteria) that is 100\% reliable in separating members from non-member
contaminants.  Many have been considered in the literature (e.g.,
Allen \etal\ 2004, Rebull \etal\ 2007, Harvey \etal\ 2007, Gutermuth
\etal\ 2008, 2009, Rebull \etal\ 2010, Rebull \etal\ 2011). Some make
use of just MIPS bands, some make use of just IRAC bands, most use a
series of color criteria, and where possible, they make use of
(sometimes substantial) ancillary data. In our case of the CG4+Sa101
region, we have some ancillary data, but the bulk of the data are
IRAC+2MASS data.  In this case, the best choice for selecting YSO
candidates is the approach developed by Gutermuth \etal\ (2008, 2009)
and adapted by Guieu \etal\ (2009, 2010) for the case in which no
extinction map is available.  This selection method starts from the
set of objects detected at all four IRAC bands and uses 2MASS and MIPS
data where possible. It implements a series of color cuts to attempt
to remove contaminants such as background galaxies and knots of
nebulosity. 

\begin{figure*}[tbp]
\epsscale{1}
\plotone{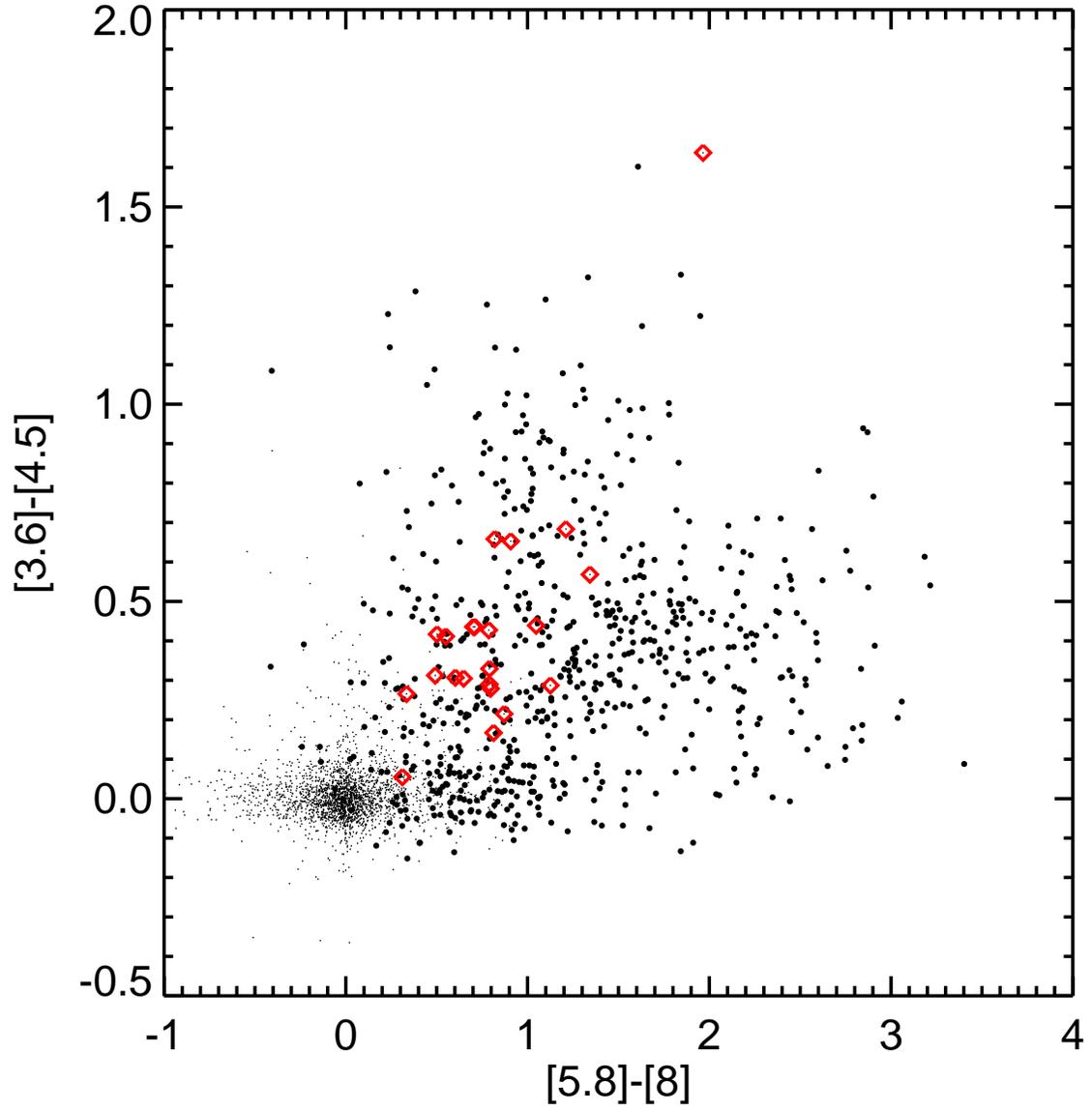}
\caption{[3.6]$-$[4.5] vs.\ [5.8]$-$[8] color-color diagram for
CG4+Sa101. Small dots are objects in the catalog; larger dots are
objects identified as contaminants, and large red diamonds highlight
our YSO candidates.  All of our Spitzer-selected YSO candidates have
colors in this diagram consistent with known YSOs, but many
contaminants do too. }
\label{fig:iracysos}
\end{figure*}

When we impose these IRAC-based color cuts, we find 25 potential YSO
candidates. We then inspected each of these in all available images
and color-color and color-magnitude diagrams. On the basis of this
inspection, we dropped three of the 25 potential YSO candidates off of
our list, though additional data will be needed to be sure that these
objects are extragalactic. Two of those dropped objects
(073542.2-470126 and 073548.5-470727) have no available data other
than IRAC, their SEDs are very flat, and they are located near the
edges of our images, far from other YSOs and nebulosity.  We suspect
that these are extragalactic contaminants.  The third object,
073243.5-464941, is returned by the IRAC selection as having a small
excess at 5.8 and 8 \mum. It is seen at 2MASS and IRAC bands, but is
undetected at 24 \mum\ to a fairly stringent limit (971 $\mu$Jy, or
9.67 mag). If it has an excess at  24 \mum, it is a very small
excess.  The 70 \mum\ data do not cover this object, so there are no
constraints (not even limits) at 70 \mum. Moreover, it is a relatively
faint source next to a very bright source, located far from any
nebulosity.  The wings of the  bright source are likely to adversely
affect the photometric accuracy of the measurements associated with
this object.  Because of this uncertainty and the very low excess as
measured, we have dropped this object from our YSO candidate list as a
likely foreground or background star.

The remaining 22 YSO candidates that pass the color cuts are shown in
Figure~\ref{fig:iracysos}. In this Figure, objects with zero color are
likely foreground or background stars (photospheres without disks),
though some could be young stars that have already shed their disks.
Young stars with circumstellar disks are generally red in both IRAC
colors, but contaminants such as galaxies also may have these colors.
All of the 22 objects highlighted in this Figure have IRAC colors
consistent with young stars with disks.

\begin{figure*}[tbp]
\epsscale{1}
\plotone{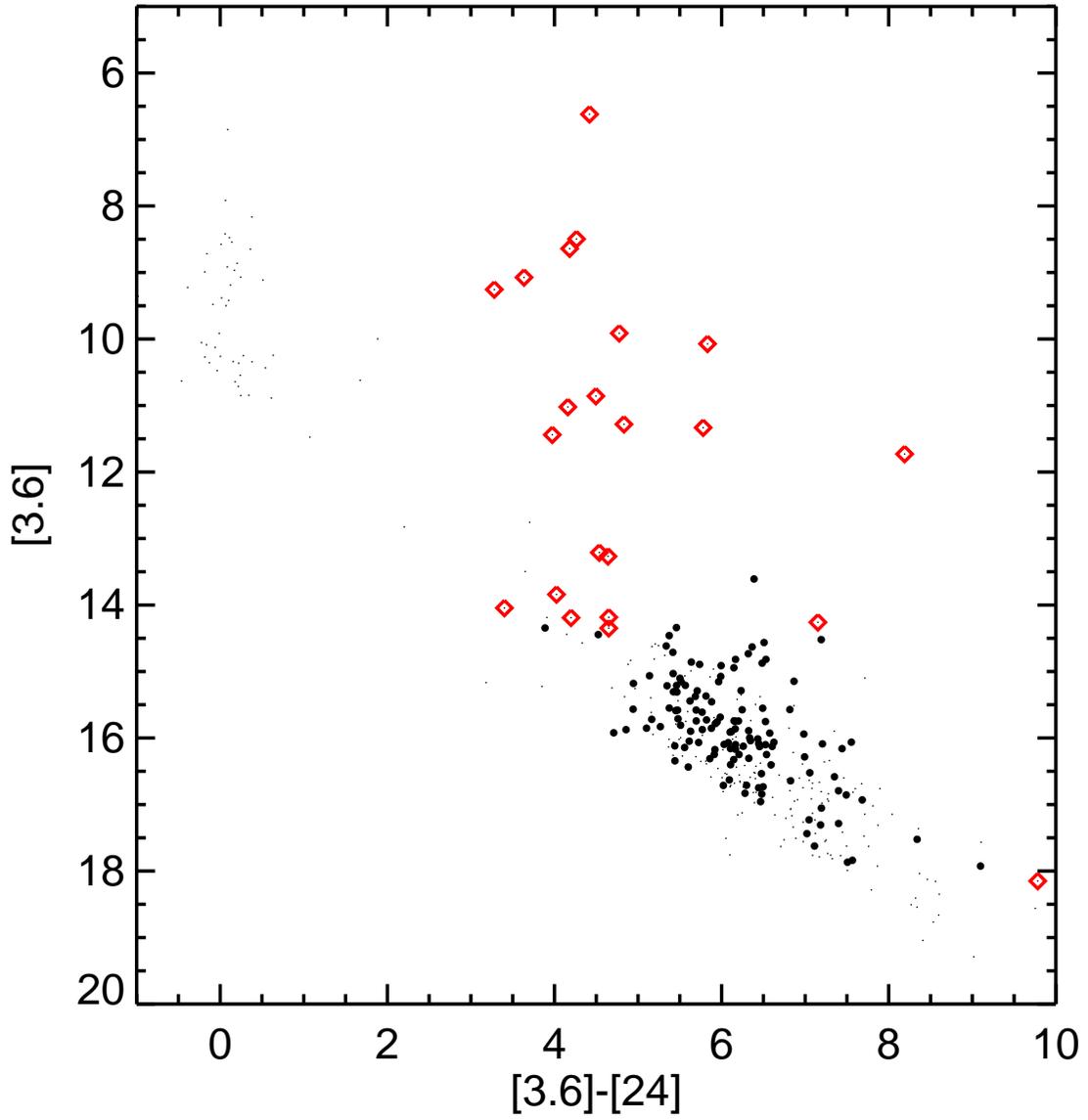}
\caption{[3.6] vs.\ [3.6]$-$[24] color-magnitude diagram for
CG4+Sa101. Small dots are objects in the catalog; larger dots are
objects identified as contaminants, and large red diamonds highlight
our IRAC-selected YSO candidates.  All of the IRAC-selected
YSO candidates have colors in this diagram consistent with known YSOs.
}
\label{fig:3324ysos}
\end{figure*}

\begin{figure*}[tbp]
\epsscale{1}
\plotone{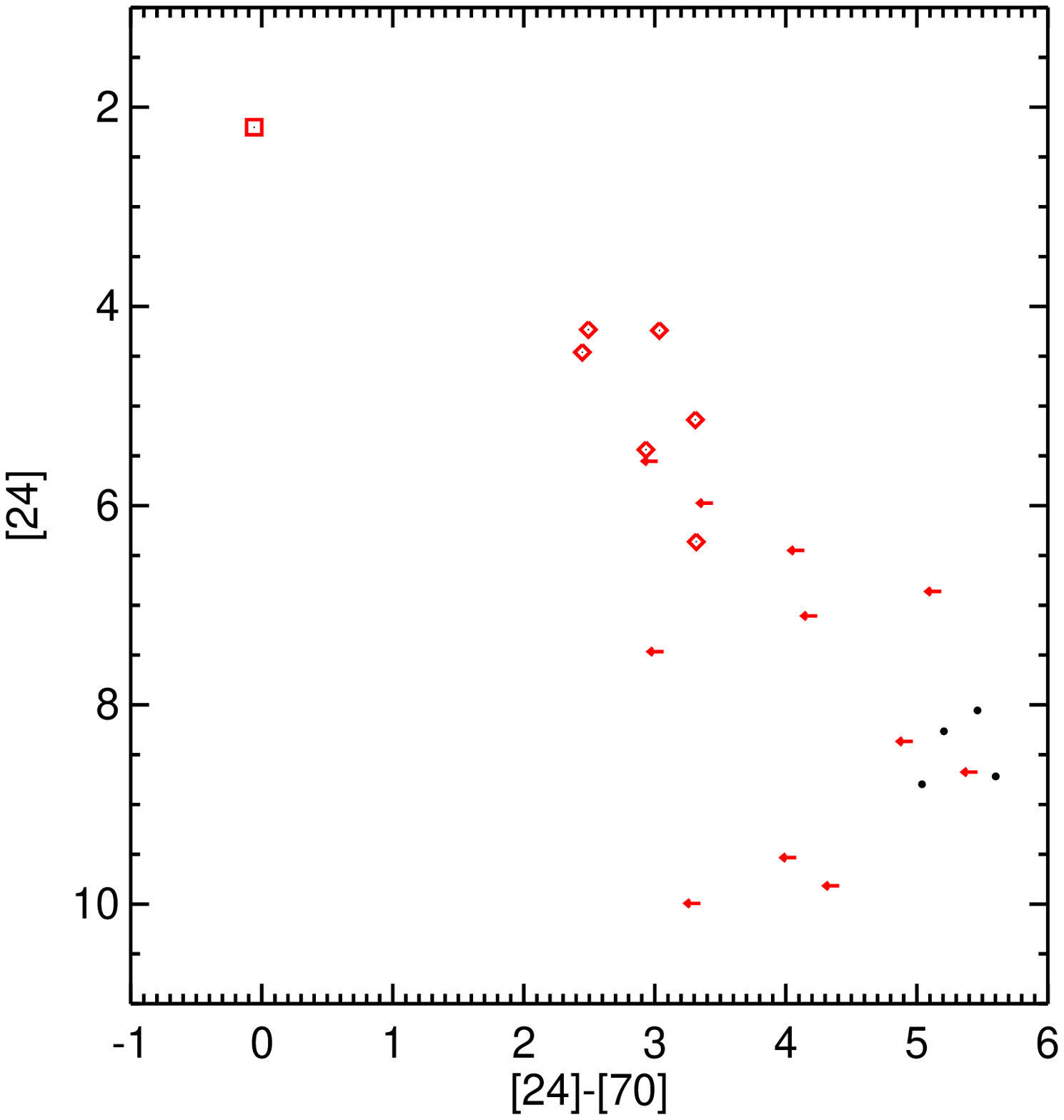}
\caption{[24] vs.\ [24]$-$[70] color-magnitude diagram for CG4+Sa101.
As in prior plots, dots are objects identified as contaminants, and
large red diamonds highlight our IRAC-selected YSO candidates. A red
square highlights a very bright likely background star; see the text.
The arrows indicate the positions of the YSO candidates for which we
could obtain upper limits at 70 \mum. The IRAC-selected YSO candidates
have colors in this diagram consistent with known YSOs. }
\label{fig:2470ysos}
\end{figure*}

The Gutermuth \etal\ (2008, 2009) selection criteria have provisions
for adding stars to the list of candidate YSOs based on properties at
other bands, such as MIPS bands.  We now investigate the properties of
objects in our catalogs at the MIPS bands to see if we should add
additional objects to our list of YSO candidates. In summary of the
rest of this section, while we find some interesting objects, in the
end, we do not add any more YSO candidates to our list.

Young stars having inner disk holes and thus excesses at only the
longest bands can be revealed via comparison of the 24 \mum\
measurement to a shorter band, such as \ks\ or [3.6].  If the data are
available, one should use [3.6] vs.\ [3.6]$-$[24] rather than $K_s$ vs.\
$K_s-[24]$.  There is an intrinsic spread in $K_s-[24]$ photospheric
colors that is not present in [3.6]$-$[24] because late type stars are
not colorless at $K_s-[24]$ (Gautier \etal\ 2007).  The effects of
reddening are stronger at \ks\ than at 3.6 \mum. And, if 2MASS is the
only source of \ks, even short 3.6 \mum\ integrations can reach
fainter sources than 2MASS does.  In our case of CG4+Sa101, the IRAC
coverage is larger than the MIPS coverage, and so we use [3.6] vs.\
[3.6]$-$[24] to look for any objects with an excess starting at 24
\mum.  

Figure~\ref{fig:3324ysos} shows this [3.6] vs.\ [3.6]$-$[24] diagram,
with the same notation as the prior figure. Ordinary stellar
photospheres (likely foreground or background stars) have
$[3.6]-[24]\sim$0, and galaxies make up the large, elongated source
concentration near [3.6]$-$[24]$\sim$6, [3.6]$\sim$16.  Objects not in
this region, e.g., the brighter and/or redder objects, are less likely
to be part of the Galactic or extragalactic backgrounds, and more
likely to be YSOs with a 24 \mum\ excess.  Most of the IRAC-identified
YSO candidates are indeed in the region of this diagram occupied by
other known YSOs (see, e.g., Rebull \etal\ 2010, 2011, Guieu \etal\
2009, 2010). There is one (073049.1-470209) that is among the reddest
objects in this diagram, near [3.6]$-$[24]$\sim$10; see
Appendix~\ref{sec:073049.1-470209} for more on this specific object.  
Most of the objects already ruled out as YSO candidates based on their
IRAC properties are in the extragalactic concentration of sources. The
objects with $[3.6]-[24]\sim$0 do not have apparent excesses, but
there are eight additional objects with [3.6]$-$[24]$>$1 and
[3.6]$<$14.5 that seem to have the right placement in this diagram to
be YSO candidates.  We investigated each of these candidates, and none
had evidence based on SED shape, significance of excess, or appearance
in the images compelling enough to have us add them to our list of YSO
candidates. The apparent small excesses just at 24 \mum\ are most
likely due to source confusion at the lower resolution 24 \mum\ band,
with either a background source or a low-mass companion.  The most
compelling one based on the SED is 073355.0-464838, but the source
seems to be confused with a nearby source that emerges at 8 \mum, and
we strongly suspect that the 24 \mum\ flux instead corresponds to the
object appearing at 8 \mum, rather than the point source seen at 8
\mum\  and shorter bands.  We do not add this source to our list of
YSO candidates at this time. Higher spatial resolution 24 \mum\
observations would be required to resolve this issue.

Some of the brightest stars in the CG4+Sa101 region are saturated in
at least the first two IRAC bands, so neither of the YSO search
mechanisms we have used thus far would find them.  However, these
sources could be YSOs, and they are not all saturated in MIPS bands. 
As our last attempt to search for YSO candidates with infrared
excesses, Figure~\ref{fig:2470ysos} shows the [24] vs.\ [24]$-$[70]
diagram for our region.  This diagram for this region is sparse, but a
better-populated diagram (see, e.g., Rebull \etal\ 2010) would
basically resemble Figure~\ref{fig:3324ysos}, with photospheres being
bright and colorless (having [24]$-$[70]$\sim$0), and galaxies red and
faint. In Figure~\ref{fig:2470ysos}, the few points that are available
include things that are galaxies (things that have been ruled out as
contaminants based on IRAC) which are in the extragalactic part of the
diagram here as well, six detections that have colors consistent with
YSOs, many limits for our YSO candidates, and one thing (square in
Figure~\ref{fig:2470ysos}) that is too bright to be a likely galaxy,
and does not appear to be particularly red in this diagram.  This very
bright object is 073339.7-464839, and its SED suggests at first glance
that it might be a YSO with a small excess just at 24 and 70 \mum. It
is detected at $JHK_s$, [5.8], and [8], and the $K_s$, [5.8], and [8]
measurements are all consistent with one Rayleigh-Jeans tail, and the
[24] and [70] measurements are offset on a different, redder
Rayleigh-Jeans tail, as might be consistent with a small
thermally-heated dust disk. However, it is a very bright object, and
saturated at the shortest two IRAC bands; the photometry at
$JHK_s$[5.8][8] may also be compromised beyond what our formal errors
suggest. It is matched in Simbad to IRAS 07321-4642. We suspect that
this is a background asymptotic giant branch (AGB) star, or another
sort of bright background giant, and not a legitimate YSO candidate.
We do not include it on our YSO candidate list.

We move ahead from here with the 22 IRAC-selected YSO candidates, and
now investigate their multi-band properties.

\section{Properties of selected YSO candidates}
\label{sec:properties}

\subsection{Optical properties}

\begin{figure*}[tbp]
\epsscale{1}
\plotone{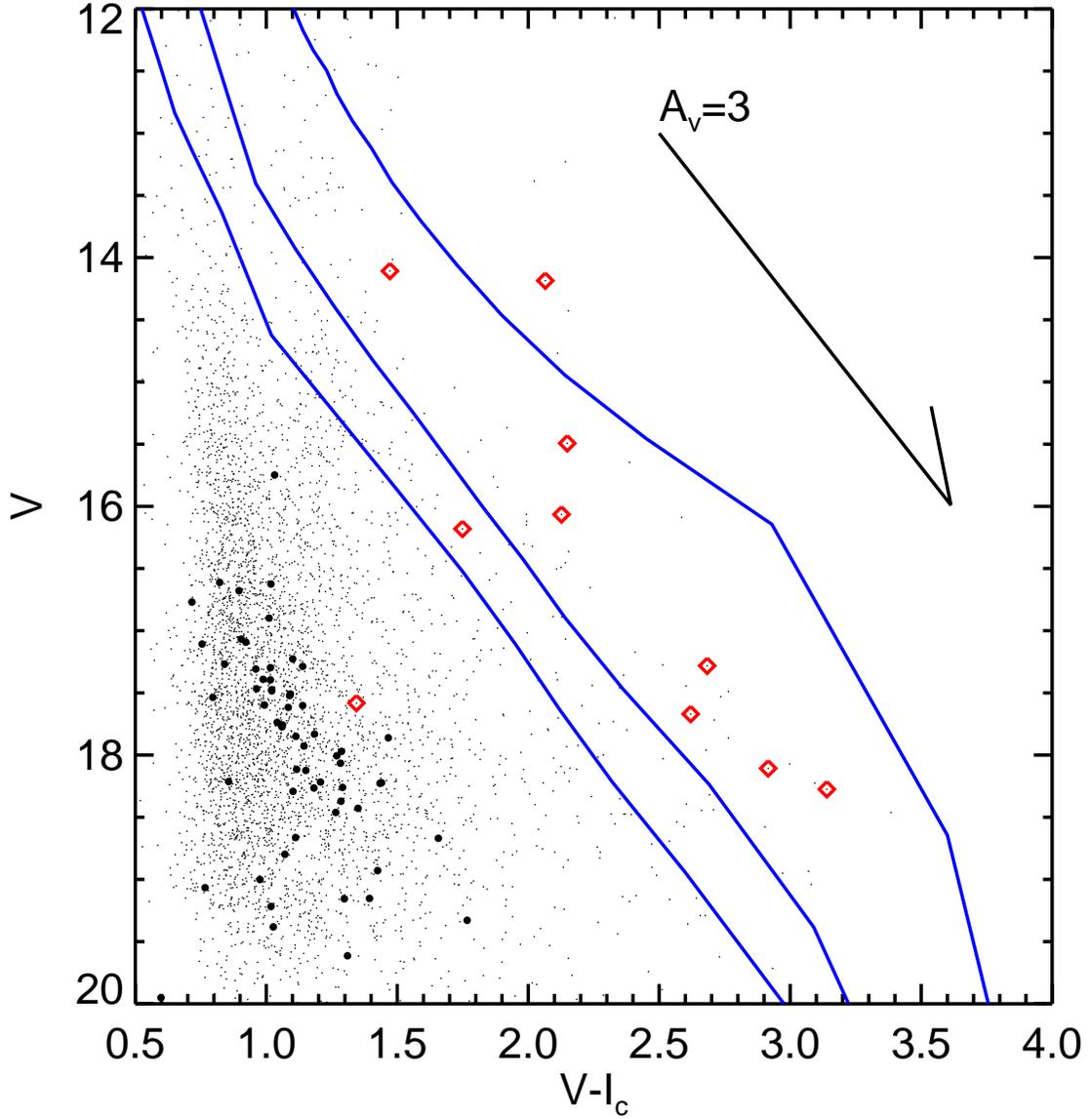}
\caption{$V$ vs.\ $(V-I_c)$ color-magnitude diagram for CG4+Sa101.
Small dots are objects in the catalog, larger dots are objects
identified as contaminants, and large red diamonds highlight
our YSO candidates.   Isochrones given are models from Siess
\etal\ (2000) at 1, 10, and 30 Myr, scaled to 500 pc, where we have
tuned the color-effective temperature relation such that the 100 Myr
isochrone matches that of the Pleiades single-star sequence (Stauffer
\etal\ 2007, Jeffries \etal\ 2007). A reddening vector is also
indicated.  All of the YSO candidates shown here, except for one, are
in the region occupied by young stars at 300-500 pc.}
\label{fig:vviysos}
\end{figure*}

Optical data can greatly aid in confirming or refuting YSO candidacy
because they provide constraints on the Wien side of the SED. In Guieu
\etal\ (2010), most of the IRAC-selected candidates in IC~2118 proved
to be too faint, most vividly in the optical, to be likely cluster
members. Just ten of our candidate YSOs have optical data available,
and they appear in Figure~\ref{fig:vviysos}. The objects with optical
data that have already been ruled out as YSOs based on their IRAC
properties are all well below the Siess \etal\ (2000) 30 Myr isochrone
scaled to 500 pc. One YSO candidate object appears below the 30 Myr
isochrone; 073337.6-464246 is within the distribution of clear non-YSO
points. We do not remove this object from our list, since a variety of
reasons (such as scattered light) could result in a YSO appearing
below the 30 Myr isochone; for more discussion of this object, please
see Appendix~\ref{app:073337.6-464246}. As noted above, the distance
to this association is uncertain; if the isochrones are instead scaled
to 300 pc, then one other object (073121.8-465745, a previously
identified YSO) appears to be just below the 30 Myr isochone instead
of just above it.

Deeper optical data are desirable in order to obtain magnitude
estimates for the remaining YSO candidates.

\subsection{Near-IR properties}


\begin{figure*}[tbp]
\epsscale{1}
\plotone{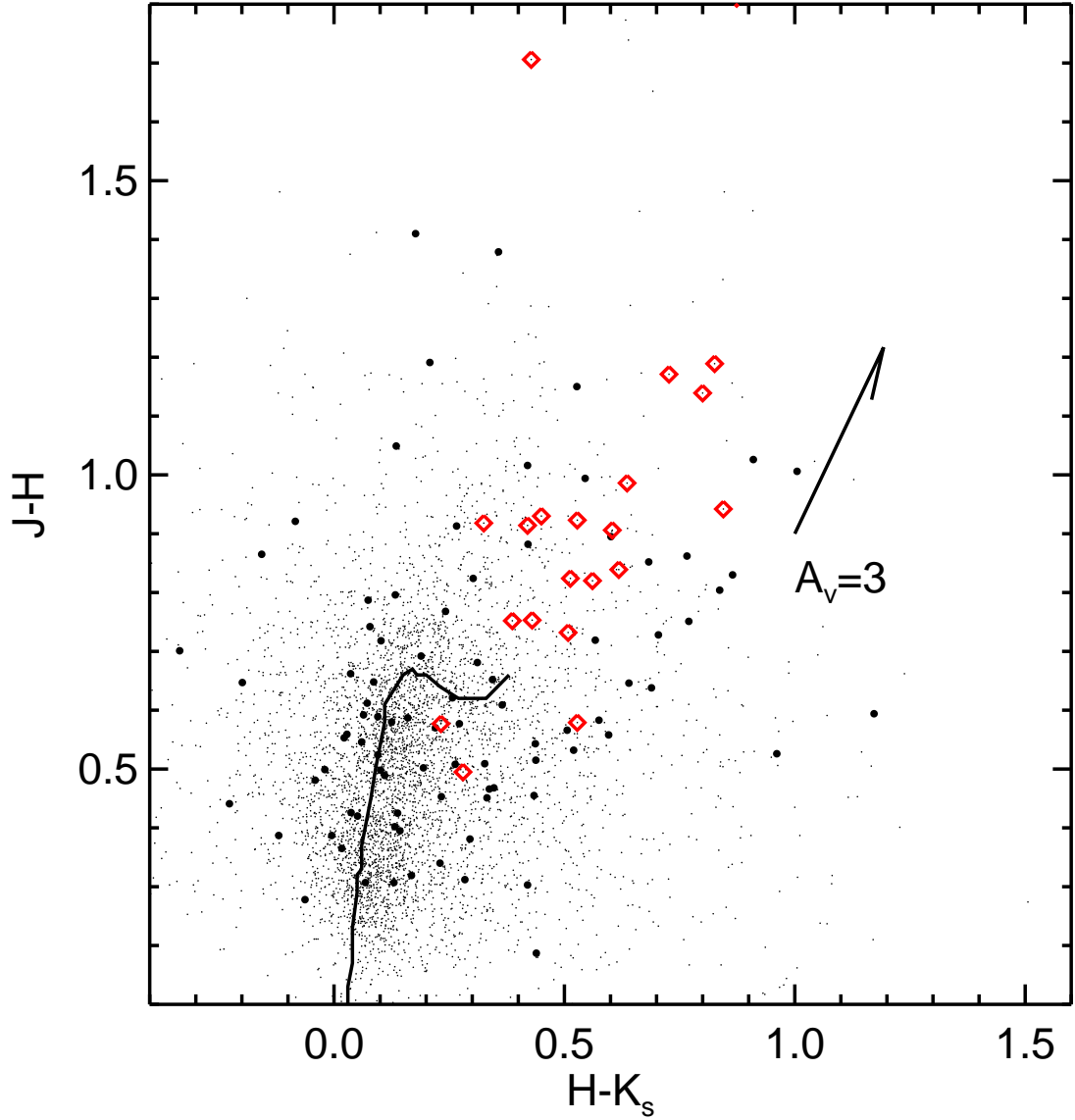}
\caption{$J-H$ vs.\ $H-K_s$ diagram for the sample, with the same
notation as earlier figures. The main sequence is indicated by a solid
line. Most of the YSO candidates have an infrared excess starting at
H-band with moderate reddening. }
\label{fig:jhkysos}
\end{figure*}


Near-IR data can also aid in confirming or refuting YSO candidacy.
Since we do not have spectral types for most of our sources, it is
difficult to estimate the degree of reddening. Figure~\ref{fig:jhkysos}
shows $J-H$ vs.\ $H-K_s$ for the sample.  This plot suggests that most
of our YSO candidates have an infrared excess with a moderate degree
of reddening.

\subsection{$B$-band properties}

\begin{figure*}[tbp]
\epsscale{1}
\plotone{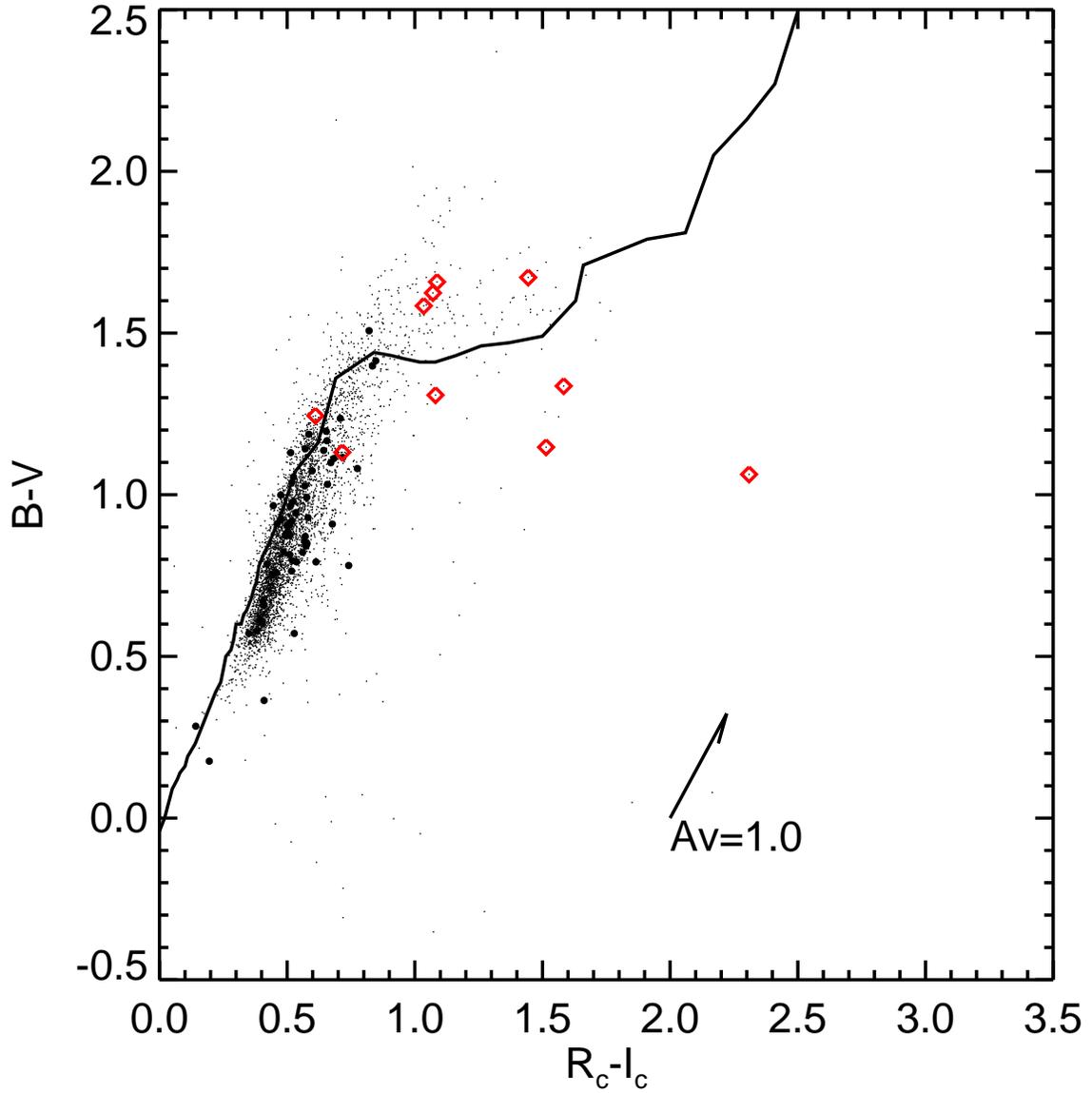}
\caption{$B-V$ vs.\ $R_c-I_c$ for the sample, with the same notation as
earlier figures. The main sequence is indicated by a solid line. Stars
with a $B$-band excess and relatively small values of \av\ would be
blue, e.g., below the line in this figure. Objects above the line on
the upper right of this figure are pushed into that location by high
\av.  At least 4 and probably 8 of the YSO candidates have a $B$
excess, most likely from mass accretion. }
\label{fig:uvysos}
\end{figure*}

Young stars that are actively accreting from their circumstellar disk
can have excess ultraviolet emission at $U$ or $B$ bands, or even
longer bands during periods of intense accretion. However, these bands
are also most sensitive to reddening.  Figure~\ref{fig:uvysos} shows
$B-V$ vs.\ $R_c-I_c$ for the sample, with the main sequence indicated as
a solid line.  Stars with a $B$-band excess and relatively small
values of \av\ would be blue, e.g., below the line in this figure.
Objects above the line on the upper right of this figure are pushed
into that location by high \av.  At least 4 of the YSO candidates have
a $B$ excess, most likely from mass accretion; four more appear to
have been pushed from the region of clear $B$ excess by high \av.
Similar results for the same objects are obtained from the $B-V$ vs.\
$V-I_c$ plot. The individual objects are listed in the Appendix.

\subsection{Spectral Energy Distributions}

\begin{deluxetable}{cccccccccccccc}
\tablecaption{Multiband measurements of Spitzer-identified YSO candidates in the CG4+Sa101
region\label{tab:ourysos}}
\rotate
\tabletypesize{\tiny}
\tablewidth{0pt}
\tablehead{
\colhead{name}  & 
\colhead{$B$ (mag)}  &
\colhead{$V$ (mag)} & \colhead{$R_c$ (mag)} & \colhead{$I_c$ (mag)} & 
\colhead{$J$ (mag)} &  \colhead{$H$ (mag)} &\colhead{$K_s$ (mag)} & 
\colhead{[3.6] (mag)} & \colhead{[4.5] (mag)} &
\colhead{[5.8] (mag)} & \colhead{[8.0] (mag)} & \colhead{[24]
(mag)}  & \colhead{[70] (mag)} }
\startdata
   073049.1-470209&            $\ldots$&            $\ldots$&            $\ldots$&            $\ldots$&            $\ldots$&            $\ldots$&            $\ldots$&   18.15$\pm$   0.28&   16.51$\pm$   0.17&   14.54$\pm$   0.11&   12.57$\pm$   0.08&    8.37$\pm$   0.05&     $>$        3.40\\
   073049.8-465806&            $\ldots$&            $\ldots$&            $\ldots$&            $\ldots$&     $>$       17.32&     $>$       15.61&   15.19$\pm$   0.17&   14.26$\pm$   0.08&   13.69$\pm$   0.14&   12.39$\pm$   0.08&   11.05$\pm$   0.07&    7.11$\pm$   0.05&     $>$        2.87\\
   073053.6-465742&            $\ldots$&            $\ldots$&            $\ldots$&            $\ldots$&   16.89$\pm$   0.18&   15.70$\pm$   0.10&   14.87$\pm$   0.11&   14.18$\pm$   0.08&   13.87$\pm$   0.09&   13.53$\pm$   0.09&   13.04$\pm$   0.09&    9.53$\pm$   0.09&     $>$        5.45\\
   073057.5-465611&   19.61$\pm$   0.06&   18.27$\pm$   0.06&   16.72$\pm$   0.06&   15.13$\pm$   0.06&   12.86$\pm$   0.02&   11.93$\pm$   0.02&   11.40$\pm$   0.02&   10.86$\pm$   0.07&   10.43$\pm$   0.07&   10.04$\pm$   0.07&    9.26$\pm$   0.07&    6.36$\pm$   0.05&    3.05$\pm$   0.22\\
   073106.5-465454&            $\ldots$&            $\ldots$&            $\ldots$&            $\ldots$&   16.72$\pm$   0.14&   14.90$\pm$   0.05&   13.65$\pm$   0.04&   11.28$\pm$   0.07&   10.63$\pm$   0.07&   10.00$\pm$   0.07&    9.09$\pm$   0.07&    6.45$\pm$   0.04&     $>$        2.31\\
   073108.4-470130&            $\ldots$&            $\ldots$&            $\ldots$&            $\ldots$&   15.21$\pm$   0.04&   14.39$\pm$   0.03&   13.83$\pm$   0.05&   13.21$\pm$   0.08&   12.88$\pm$   0.08&   12.50$\pm$   0.08&   11.71$\pm$   0.08&    8.68$\pm$   0.05&     $>$        3.21\\
   073109.9-465750&            $\ldots$&            $\ldots$&            $\ldots$&            $\ldots$&   16.55$\pm$   0.12&   15.71$\pm$   0.13&   15.09$\pm$   0.15&   14.35$\pm$   0.09&   13.91$\pm$   0.09&   13.52$\pm$   0.09&   12.82$\pm$   0.08&    9.70$\pm$   0.08&     $>$        3.57\\
   073110.8-470032&   17.12$\pm$   0.03&   15.49$\pm$   0.03&   14.41$\pm$   0.03&   13.34$\pm$   0.03&   11.20$\pm$   0.02&   10.22$\pm$   0.02&    9.58$\pm$   0.02&    8.64$\pm$   0.05&    8.22$\pm$   0.05&    7.88$\pm$   0.07&    7.38$\pm$   0.05&    4.46$\pm$   0.04&    2.01$\pm$   0.22\\
   073114.6-465842&            $\ldots$&            $\ldots$&            $\ldots$&            $\ldots$&   13.56$\pm$   0.03&   11.75$\pm$   0.03&   10.88$\pm$   0.02&   10.07$\pm$   0.08&    9.63$\pm$   0.07&    8.93$\pm$   0.07&    7.88$\pm$   0.07&    4.24$\pm$   0.04&    1.21$\pm$   0.22\\
   073114.9-470055&            $\ldots$&            $\ldots$&            $\ldots$&            $\ldots$&   15.53$\pm$   0.06&   14.95$\pm$   0.05&   14.43$\pm$   0.09&   13.84$\pm$   0.08&   13.54$\pm$   0.08&   13.20$\pm$   0.08&   12.55$\pm$   0.08&    9.82$\pm$   0.09&     $>$        5.41\\
   073121.8-465745&   17.49$\pm$   0.05&   16.18$\pm$   0.05&   15.51$\pm$   0.05&   14.43$\pm$   0.05&   11.42$\pm$   0.03&   10.67$\pm$0.2\tablenotemark{a}&   10.24$\pm$0.2\tablenotemark{a}&    9.26$\pm$   0.07&    8.84$\pm$   0.07&    8.73$\pm$   0.08&    8.18$\pm$   0.07&    5.97$\pm$   0.04&     $>$        2.53\\
   073136.6-470013&   17.65$\pm$   0.04&   16.07$\pm$   0.04&   14.97$\pm$   0.04&   13.94$\pm$   0.04&   11.96$\pm$   0.02&   10.82$\pm$   0.03&   10.02$\pm$   0.02&    9.07$\pm$   0.08&    8.81$\pm$   0.05&    8.53$\pm$   0.05&    8.19$\pm$   0.05&    5.44$\pm$   0.04&    2.51$\pm$   0.22\\
   073137.4-470021&   15.84$\pm$   0.04&   14.19$\pm$   0.04&   13.21$\pm$   0.04&   12.12$\pm$   0.04&   10.45$\pm$   0.02&    9.53$\pm$   0.02&    9.11$\pm$   0.02&    8.50$\pm$   0.08&    8.21$\pm$   0.05&    7.98$\pm$   0.05&    7.19$\pm$   0.05&    4.23$\pm$   0.04&    1.74$\pm$   0.22\\
   073143.8-465818&   19.25$\pm$   0.05&   18.11$\pm$   0.05&   16.70$\pm$   0.05&   15.19$\pm$   0.05&   13.40$\pm$   0.03&   12.57$\pm$   0.02&   12.06$\pm$   0.02&   11.33$\pm$   0.05&   10.65$\pm$   0.05&    9.98$\pm$   0.05&    8.77$\pm$   0.05&    5.55$\pm$   0.04&     $>$        2.53\\
   073144.1-470008&   18.73$\pm$   0.05&   17.67$\pm$   0.05&   17.36$\pm$   0.05&   15.05$\pm$   0.05&   13.39$\pm$   0.05&   12.48$\pm$   0.07&   11.88$\pm$   0.03&   11.02$\pm$   0.05&   10.59$\pm$   0.05&   10.21$\pm$   0.05&    9.50$\pm$   0.05&    6.86$\pm$   0.04&     $>$        1.68\\
   073145.6-465917&   18.95$\pm$   0.05&   17.28$\pm$   0.05&   16.04$\pm$   0.05&   14.60$\pm$   0.05&   12.96$\pm$   0.02&   12.04$\pm$   0.02&   11.71$\pm$   0.02&   11.44$\pm$   0.05&   11.27$\pm$   0.05&   11.02$\pm$   0.05&   10.21$\pm$   0.05&    7.47$\pm$   0.05&     $>$        4.40\\
   073326.8-464842&   15.24$\pm$   0.05&   14.11$\pm$   0.05&   13.35$\pm$   0.05&   12.64$\pm$   0.05&   11.49$\pm$   0.02&   10.74$\pm$   0.02&   10.35$\pm$   0.02&    9.91$\pm$   0.07&    9.70$\pm$   0.07&    9.43$\pm$   0.07&    8.56$\pm$   0.07&    5.14$\pm$   0.04&    1.83$\pm$   0.22\\
   073337.0-465455&            $\ldots$&            $\ldots$&            $\ldots$&            $\ldots$&   15.65$\pm$   0.09&   15.15$\pm$   0.11&   14.87$\pm$   0.13&   14.04$\pm$   0.08&   13.74$\pm$   0.08&   13.39$\pm$   0.09&   12.79$\pm$   0.08&   10.64$\pm$   0.09&     $>$        3.60\\
   073337.6-464246&   18.83$\pm$   0.02&   17.58$\pm$   0.02&   16.85$\pm$   0.02&   16.24$\pm$   0.02&   15.43$\pm$   0.07&   14.85$\pm$   0.10&   14.62$\pm$   0.11&   14.19$\pm$   0.08&   14.14$\pm$   0.09&   13.75$\pm$   0.09&   13.44$\pm$   0.11&    9.99$\pm$   0.12&     $>$        6.64\\
   073406.9-465805&            $\ldots$&            $\ldots$&            $\ldots$&            $\ldots$&   14.97$\pm$   0.05&   14.24$\pm$   0.05&   13.73$\pm$   0.04&   13.27$\pm$   0.08&   12.99$\pm$   0.08&   12.67$\pm$   0.08&   11.88$\pm$   0.08&    8.63$\pm$   0.04&     $>$        1.68\\
   073425.3-465409&            $\ldots$&            $\ldots$&            $\ldots$&            $\ldots$&   13.44$\pm$0.05\tablenotemark{b}&   12.51$\pm$0.05\tablenotemark{b}&   12.06$\pm$0.06\tablenotemark{b}&   11.73$\pm$   0.07&   11.44$\pm$   0.08&   10.94$\pm$   0.07&    9.82$\pm$   0.07&    3.54$\pm$   0.04&            $\ldots$\\
   073439.9-465548&            $\ldots$&            $\ldots$&            $\ldots$&            $\ldots$&   15.98$\pm$   0.12&   15.03$\pm$   0.09&   14.19$\pm$   0.07&   12.99$\pm$   0.08&   12.33$\pm$   0.08&   11.72$\pm$   0.08&   10.90$\pm$   0.07&            $\ldots$&            $\ldots$\\
\enddata
\tablenotetext{a}{Values appear in the 2MASS Point Source Catalog
with a photometric quality (ph\_qual) flag of `E',  denoting that the
goodness of fit quality was very poor, or that the photometry fit did
not converge.  We took the values as reported, with a large error bar,
as additional constraints on the source. }
\tablenotetext{b}{2MASS $JHK_s$ photometry comes from the 2MASS
extended source catalog, not the point source catalog; see discussion
in sections \ref{sec:bandmerging} and \ref{sec:extsrc}.}
\end{deluxetable}

\begin{deluxetable}{lllllll}
\tablecaption{Final list of YSO candidates in the CG4+Sa101
region\label{tab:ourysos2}}
\rotate
\tabletypesize{\tiny}
\tablewidth{0pt}
\tablehead{
\colhead{name}  & \colhead{syn.} &  \colhead{Sp.Ty.} &
\colhead{class} &\colhead{quality\tablenotemark{a}} & \colhead{region} &
\colhead{notes} }
\startdata                            
073049.1-470209&       &    &	  I  &  C  & Sa101 & reddest and faintest in [3.6] vs.\ [3.6]$-$[24]  \\
073049.8-465806&       &    &	  I  &  B  & Sa101 &  \\
073053.6-465742&       &    &	 II  &  A  & Sa101 & small excess at 8 \mum, most of the excess at 24 \mum \\
073057.5-465611& CG-Ha2&M2: &	 II  &  A+ & Sa101 & apparently lowest mass object in this list \\
073106.5-465454&       &    &  flat  &  C  & Sa101 & somewhat discontinuous SED \\
073108.4-470130&       &    &	 II  &  A  & Sa101 &  \\
073109.9-465750&       &    &	 II  &  A  & Sa101 &  \\
073110.8-470032& CG-Ha3&K7  &	 II  &  A+ & Sa101 &  \\
073114.6-465842&       &    &  flat  &  A  & Sa101 & high \av\ likely \\
073114.9-470055&       &    &	 II  &  A  & Sa101 &  \\
073121.8-465745& CG-Ha4&K7-M0&   II  &  A+ & Sa101 & \\
073136.6-470013& CG-Ha5&K2-5&	 II  &  A+ & Sa101 &  \\
073137.4-470021& CG-Ha6&K7  &	 II  &  A+ & Sa101 &  \\
073143.8-465818&       &    &  flat  &  A  & Sa101 &  \\
073144.1-470008&       &    &	 II  &  A  & Sa101 &  \\
073145.6-465917&       &    &	 II  &  A  & Sa101 &  \\
073326.8-464842& CG-Ha7&K5  &	 II  &  A+ & CG4   & inner disk hole? \\
073337.0-465455&       &    &	 II  &  B  & CG4   &  \\
073337.6-464246&       &    &	 II  &  C  & CG4   & very low in optical color-mag diagram, near edges of maps  \\
073406.9-465805&       &    &	 II  &  A  & CG4   &  \\
073425.3-465409&       &    &	  I  &  A  & CG4   &  extended in optical, NIR\\
073439.9-465548&       &    &	 II  &  C  & CG4   &  sparse SED\\
\enddata
\tablenotetext{a}{This grade is meant to indicate rough confidence in
the liklihood that the given YSO candidate is a legitimate YSO.
Previously identified YSOs are given a grade of `A+', our highest
quality YSO candidates are grade `A', our mid-grade YSO candidates are
grade `B', and our lowest-confidence YSO candidates are grade `C'. For
discussion of individual objects (and an explanation of why each
object has that grade), please see the Appendix.}
\end{deluxetable}

\begin{figure*}[tbp]
\epsscale{1}
\plotone{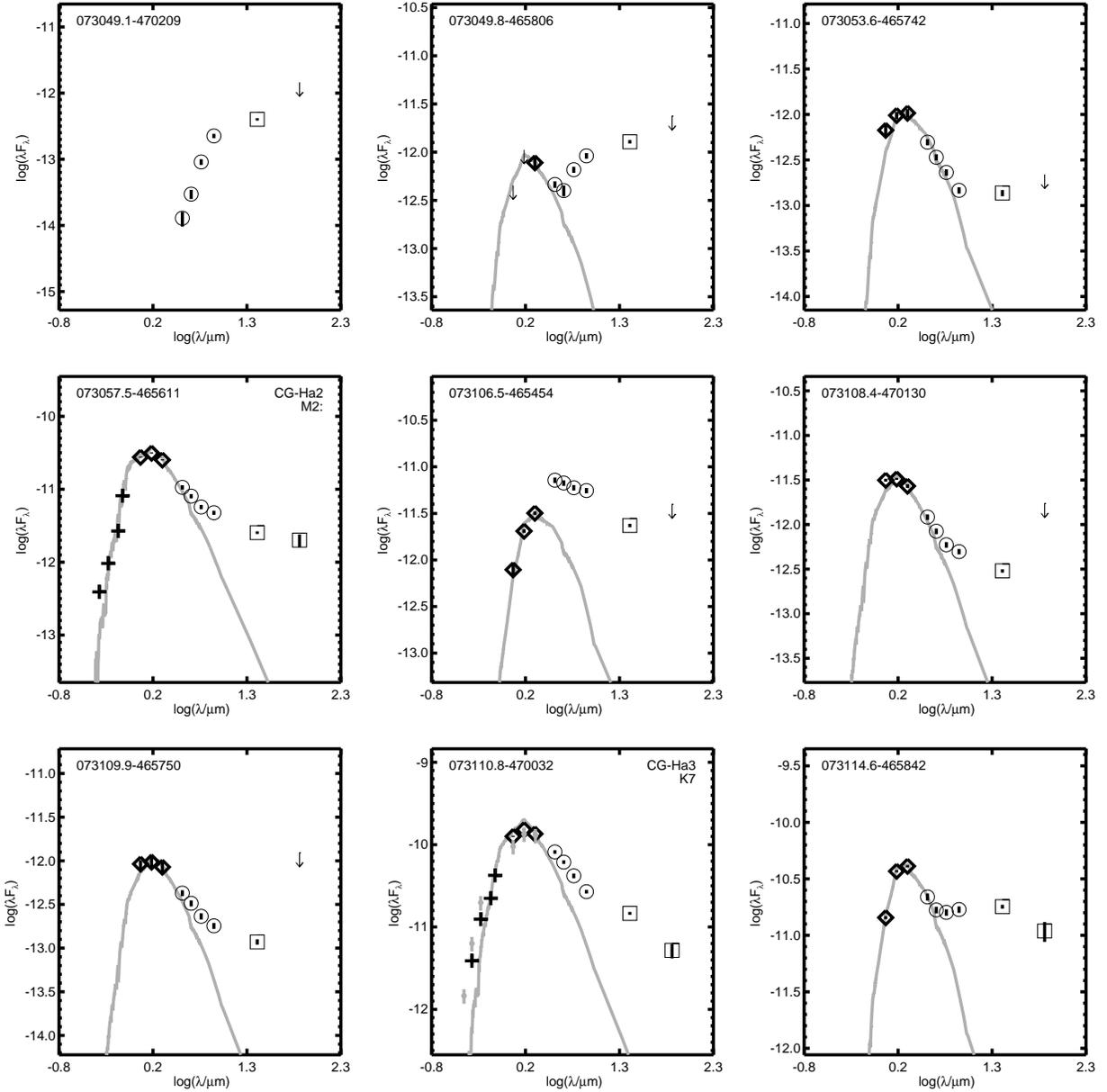}
\caption{Spectral Energy Distributions (SEDs) for the Spitzer-selected
YSOs presented here.  $+$ symbols are optical data, diamonds are 
2MASS (NIR) data, circles are IRAC data, and squares are MIPS data. 
If a previous source ID exists in Reipurth \& Pettersson (1993),  it
is in the upper right of the plot, along with their spectral type (if
it exists).  Similarly, if there are photometric data from Reipurth \&
Pettersson (1993), they are marked with light grey dots. The error bars (most
frequently far smaller than the size of the  symbol) are indicated at
the center of the symbol. Reddened models from the Kurucz-Lejeune
model grid (Lejeune \etal\ 1997, 1998) are shown for reference; see
the text. Units of $\lambda F_{\lambda}$  as presented are erg
s$^{-1}$ cm$^{-2}$, and $\lambda$ is in microns.}
\label{fig:seds1}
\end{figure*}

\begin{figure*}[tbp]
\epsscale{1}
\plotone{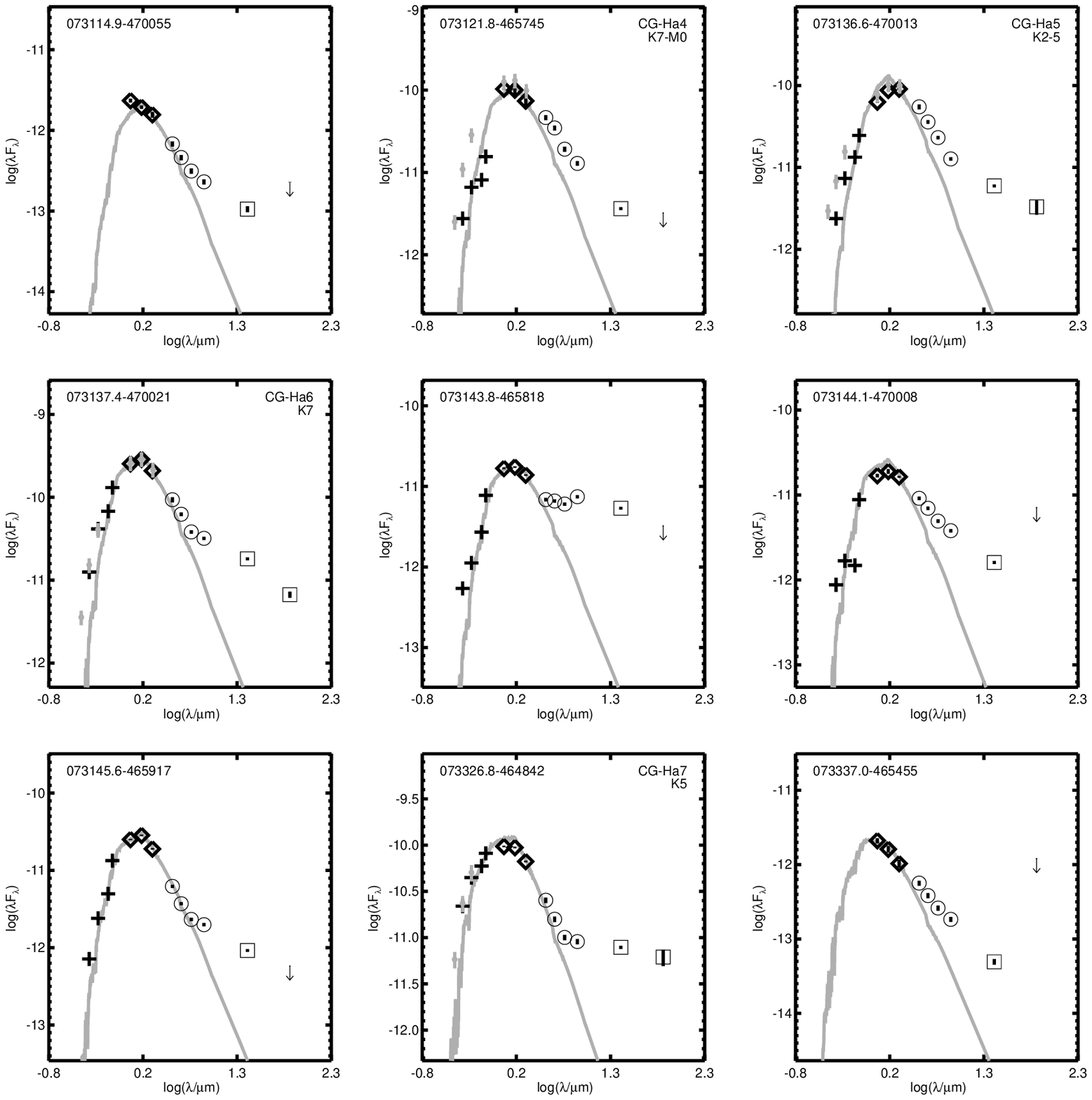}
\caption{SEDs, continued. Notation as in previous figure.}
\label{fig:seds2}
\end{figure*}

\begin{figure*}[tbp]
\epsscale{1}
\plotone{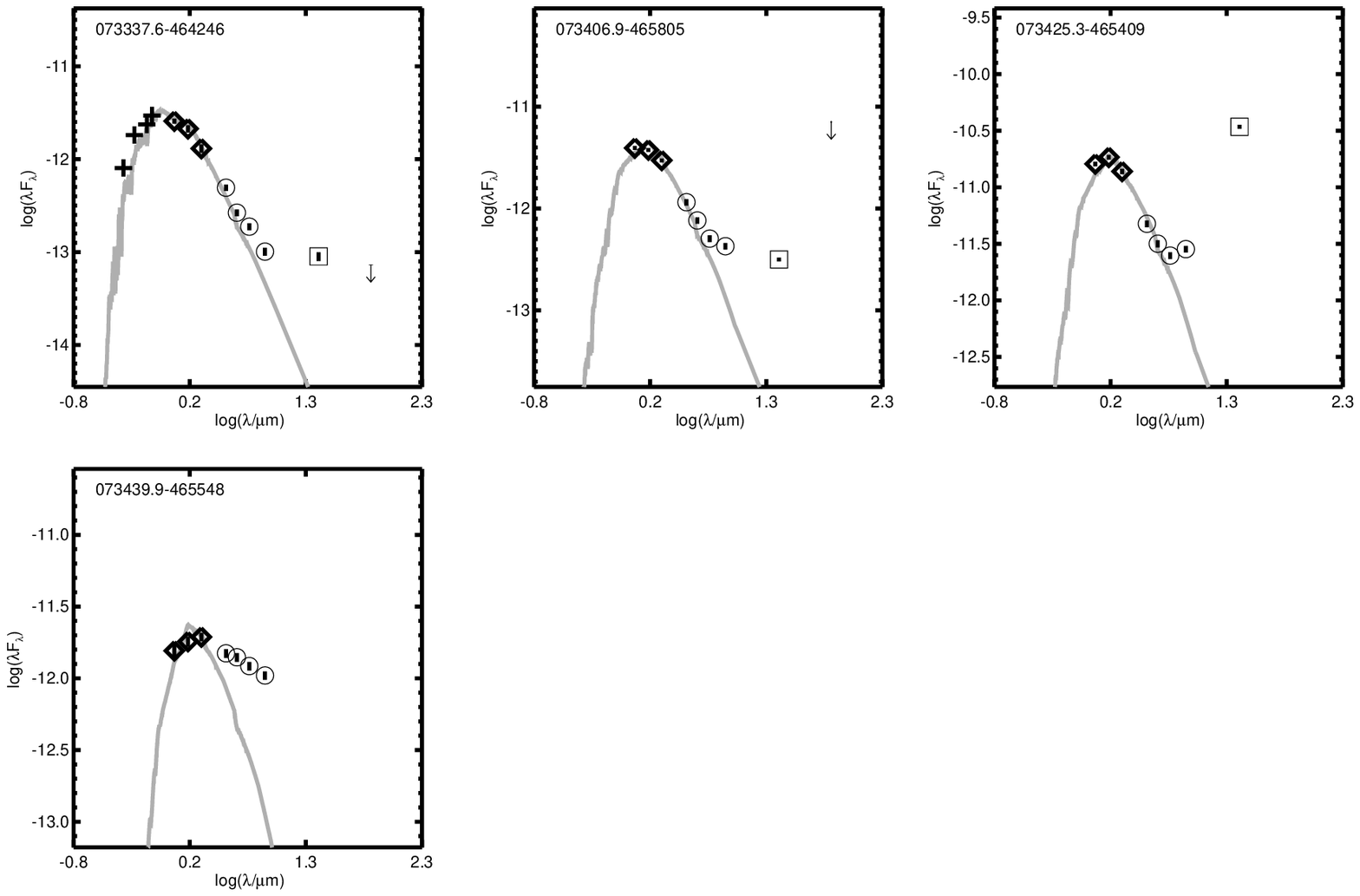}
\caption{SEDs, continued. Notation as in previous figure.}
\label{fig:seds3}
\end{figure*}

Coordinates and our measured magnitudes between $B$ and 70 \mum\ for
our 22 YSO candidates appear in Table~\ref{tab:ourysos}. Six of them
(27\%) are rediscoveries of the previously known YSOs in this region
from Table~\ref{tab:knownysos}. 
Figures~\ref{fig:seds1}--\ref{fig:seds3} are the SEDs for the YSO 
candidates.  

To guide the eye, we wished to add reddened stellar  models to the
SEDs, but spectral types are only known for the 6  previously known
YSOs.  In order to provide a reference, for the  remaining objects, we
assumed a spectral type of M0. For each object, a reddened model is
shown, selected from the Kurucz-Lejeune model grid (Lejeune \etal\
1997, 1998) and normalized to $K_s$ band  where possible (and to the
closest band otherwise).   Note that this is not meant to be a robust
fit to the object, but  rather a representative stellar SED to guide
the eye such that  the infrared excesses are immediately apparent.  In
some cases, ultraviolet excesses may also be present.  Additional
spectroscopic observations are needed to better constrain these fits.

In the spirit of Wilking \etal\ (2001), we define the near- to mid-IR
slope of the SED, $\alpha = d \log \lambda F_{\lambda}/d \log 
\lambda$,  where  $\alpha > 0.3$ for a Class I, 0.3 to $-$0.3 for a
flat-spectrum  source, $-$0.3 to $-$1.6 for a Class II, and $<-$1.6
for a Class III.  For each of the YSO candidate objects in our sample,
we performed a simple ordinary least squares linear fit to all
available photometry (just detections, not including upper or lower
limits) between 2 and 24 $\mu$m, inclusive.  Note that errors on the
infrared points are so small as to not affect the fitted SED slope. 
The precise definition of $\alpha$ can vary, resulting in different
classifications for certain objects. Classification via this method is
provided specifically to enable comparison within this paper via
internally consistent means. Note that the formal classification puts
no lower limit on the colors of Class III objects (thereby including
those with SEDs resembling bare stellar photospheres, and allowing for
other criteria to define youth).  By searching for IR excesses, we are
incomplete in our sample of Class III objects.   The classes for the
YSO (previously known and candidate) sample appear in 
Table~\ref{tab:ourysos2}. Out of the 22 stars, 16 (73\%) are Class II.

Based on the SEDs and location in several color-color and
color-magnitude diagrams, we have ranked the YSO candidates loosely
into three bins: high likelihood of being YSOs (grade A), mid-grade
quality (grade B),  and relatively low likelihood of being YSOs (grade
C). This grade also appears in Table~\ref{tab:ourysos2}. Most of them
(16) are in the grade A bin, which includes the 6
previously identified ones.  Comments on individual objects (including
justifications for the grades that were given) appear in the Appendix.

\begin{figure*}[tbp]
\epsscale{1}
\plotone{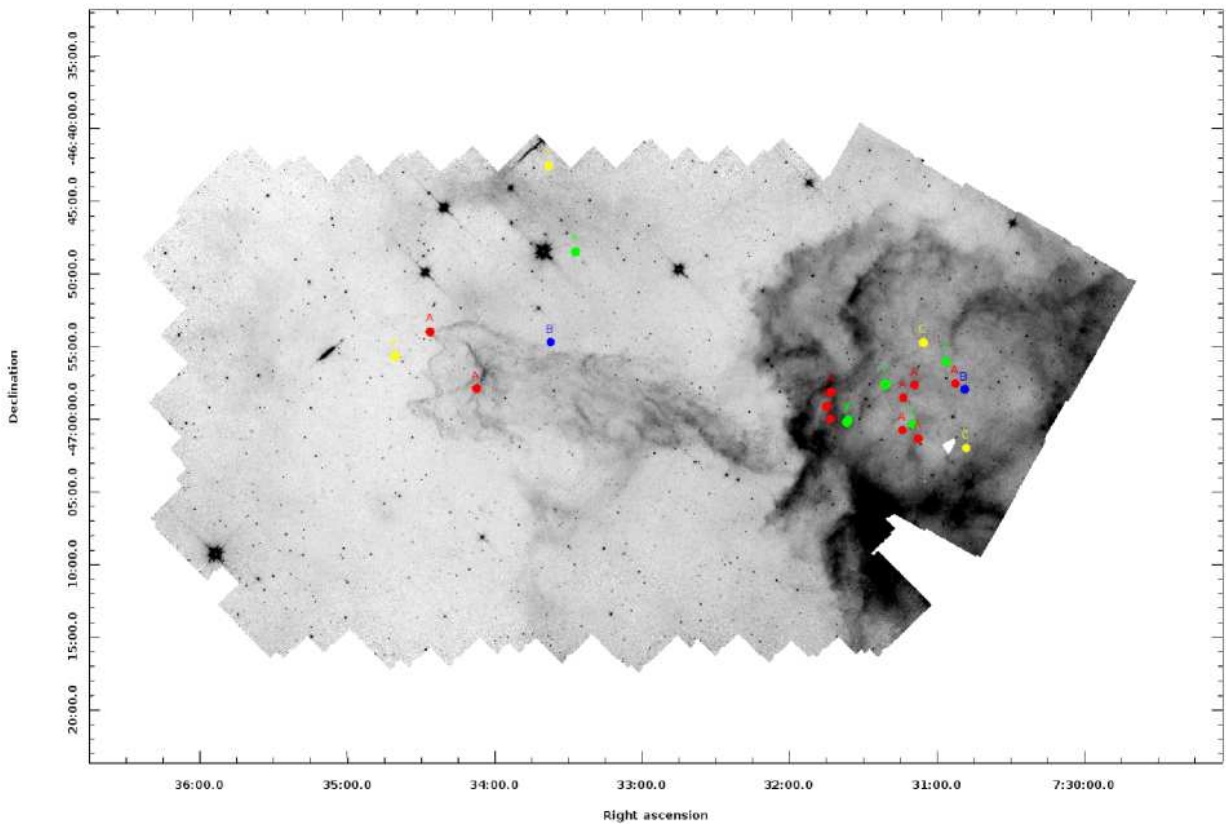}
\caption{Reverse greyscale mosaic of 8 \mum\ data with the locations
of the YSOs (previously known and new candidates) indicated, color-coded
by YSO quality grade. Previously known YSOs are indicated by a green
dot with a ``Y", high-quality (grade `A') YSOs are indicated by a red
dot and an ``A'', mid-quality (grade `B') YSOs are indicated by a blue
dot and a ``B'', and low-quality (grade `C') YSOs are indicated by a
yellow dot and a ``C.''  }
\label{fig:i4withysos}
\end{figure*}

Figure~\ref{fig:i4withysos} shows the 8 \mum\ mosaic with the 
positions of the YSO candidates overlaid, color-coded by YSO  quality.
Most of the grade A and B objects are clustered near the
previously-known YSOs. The Sa101 region has a relatively tight
clumping of most (16) of the YSOs, with a median nearest neighbor
distance of $\sim$62$\arcsec$. The CG4 region has 6 YSOs, much less
tightly clumped, with a median nearest neighbor distance nearly 5
times larger, $\sim$301$\arcsec$.  Clustering is also very commonly
found among young stars, so especially in the case of the Sa101
association, the fact that they are clustered also bolsters the case
that they are legitimate YSOs.

\begin{figure*}[tbp]
\epsscale{1}
\plotone{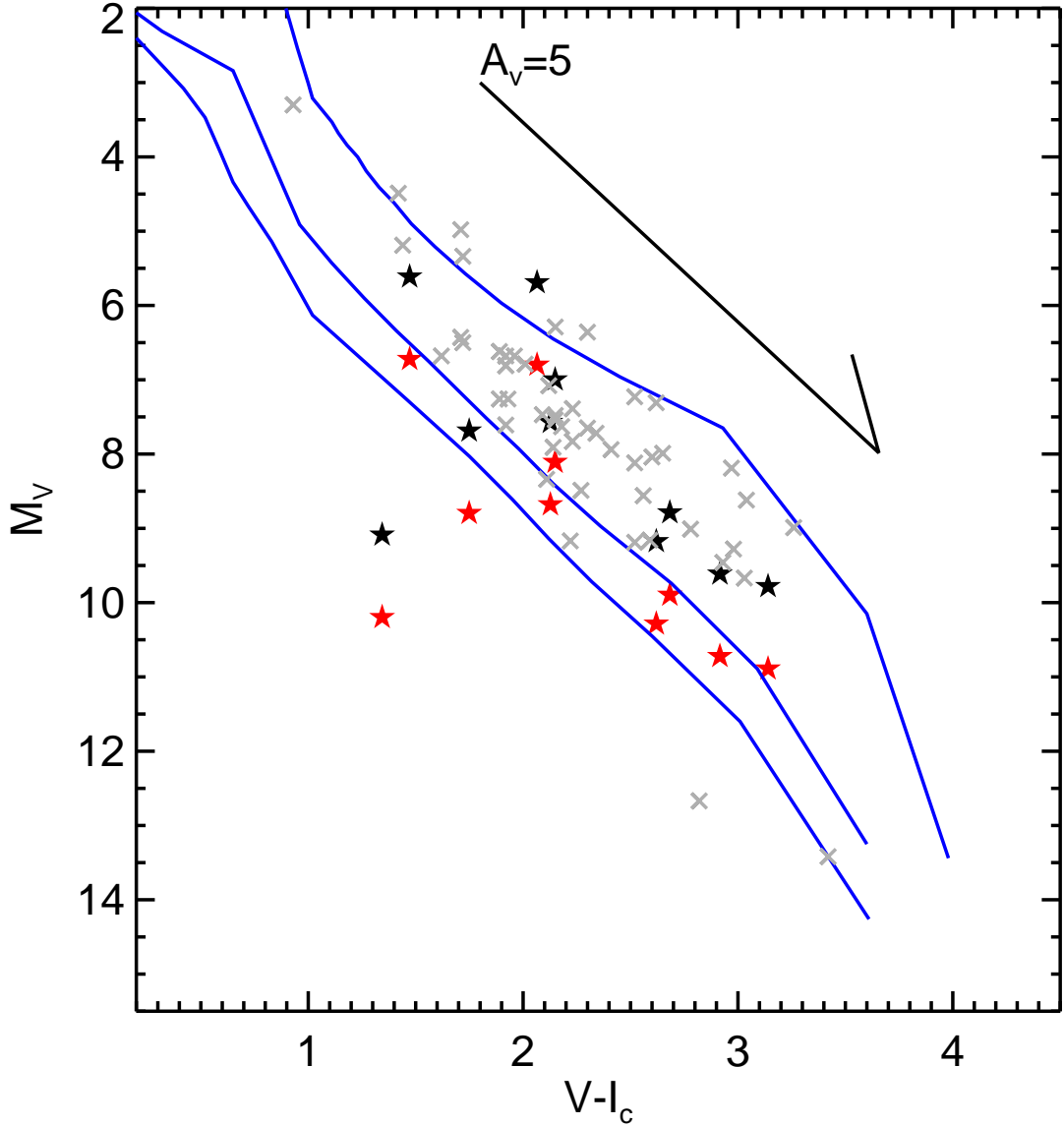}
\caption{Optical $M_V$ vs.\ $V-I_C$ color-magnitude diagram.  The
Siess \etal\ (2000) isochrones are included (1, 10, and 30 Myr), but 
shifted to absolute $M_V$.  The black stars are our YSO
candidates, assuming a distance of 
500~pc, and the red stars are our YSO candidates, assuming a
distance of 300~pc.  The grey $\times$ symbols are Taurus YSOs (from
Rebull \etal\ 2010 and G\"udel et~al.\ 2007 and references therein), 
taken to be at 140~pc; Torres \etal\ (2007, 2009).  The Taurus
distribution is broad and there are many fewer CG4+Sa101 stars, but this
distribution weakly suggests that CG4+Sa101 is farther rather than
closer (see text).  }
\label{fig:fig_mvvi}
\end{figure*}

Because the distance to this association is uncertain, we looked at
whether the relative placement of stars in the optical color-magnitude
diagram could be used to constrain the distance to the stars.
Figure~\ref{fig:fig_mvvi} presents the $M_V$ vs.\ $V-I$ CMD, comparing
CG4+Sa101 stars to YSOs in Taurus (with data from Rebull \etal\ 2010,
G\"udel et al.\ 2007, and references therein).  Based on morphological
grounds (e.g., the degree to which the YSOs are still embedded in
their natal gas), we expect that the CG4+Sa101 stars might be slightly
younger than the often more physically dispersed Taurus stars. On the
other hand, based on the ratio of Class I to Class II sources, the
CG4+Sa101 objects might be slightly older than Taurus.  In
Figure~\ref{fig:fig_mvvi}, there are not many CG4+Sa101 objects, and
the distribution is broad, but assuming a distance of 500 pc, then
CG4+Sa101 appears to be quite comparable in age to Taurus at $\sim$3
Myr. Assuming a distance of 300 pc, the CG4+Sa101 stars are on the
whole older than the Taurus stars. Figure~\ref{fig:fig_mvvi} thus
weakly suggests that CG4+Sa101 is farther rather than closer.

\section{The Galaxy}
\label{sec:galaxy}

The galaxy in our field, ESO 257-~G~019, has not been  the target of
any case studies before. The galaxy is classified as type `SB(s)cd?'
in NED, the NASA Extragalactic Database. We measure a major axis
length of 3\farcm3 or 34.7 kpc at the assumed distance of 36.1 Mpc at
the 3$\sigma$ level above the noise in our channel 1 image. We also
measured  a minor axis length of 0\farcm8 or 8.4 kpc in the same
image for this highly inclined galaxy. The surface brightness
distribution is close to exponential within the central 3 kpc. The
galaxy has a clumpy structure at larger distances from its center,
with the most prominent clumps appearing about 26\arcsec\ and
73\arcsec\ (along the major axis) to the northwest and 26\arcsec\ to 
the southeast (just outside of the plane of the galaxy). We also
measured the IRAC [3.6]$-$[4.5] and [3.6]$-$[8.0] colors by using all
the pixels above the 10$\sigma$ level. The [3.6]$-$[4.5] color is 0.0
and the [3.6]$-$[8.0] color is 1.9, both close to the typical values
for late-type galaxies, as determined by Pahre \etal\ (2004), and
consistent with the galaxy classification as given in NED. ESO
257-~G~019 appears to be a fairly isolated galaxy, with no nearby
companions within our mapped areas.

\section{Conclusions}
\label{sec:concl}

We used Spitzer Space Telescope data from the Spitzer Heritage Archive
to search for new candidate young stars in the CG4+Sa101 region of the
Gum Nebula. This region appears to be actively forming young stars,
perhaps as a result of the passage of an ionization front from the
stars powering the Gum Nebula (Reipurth \& Pettersson 1993).  We
rediscovered all six of the previously identified young stars in our
maps as having excesses at Spitzer bands. We have also discovered 16
entirely new young star candidates based on their Spitzer properties.
We used optical ground-based data and near-infrared data from 2MASS to
help constrain the SEDs of these new young star candidates. We have
sorted the new young star candidates into grades of confidence that
they are, in fact, legitimate new young stars. We find 16 high
confidence (grade ``A") objects, including the 6 previously identified
YSOs, 2 mid-grade confidence (grade ``B"), and 4 low-grade confidence
(grade ``C'') young star candidates. For all of the new young star
candidates, though, additional data will be needed, such as optical
photometry where it is missing, and optical spectroscopy to obtain
spectral types (and rule out extragalactic contamination).  Most of
the new objects are clustered in the Sa101 region, and most are SED
Class II.

\acknowledgements 

This work was performed as part of the NASA/IPAC Teacher Archive
Research Program (NITARP; http://nitarp.ipac.caltech.edu), class of
2010.  We acknowledge here all of the students and other folks who
contributed their time and energy to this work and the related poster
papers  presented at the January 2011 American Astronomical Society
(AAS) meeting in Seattle, WA. They include: With V. Hoette: C.
Gartner, J. VanDerMolen, L. Gamble, L. Matsche, A. McCartney, M.
Doering, S. Brown, R. Wernis, J. Wirth, M. Berthoud. With C. Johnson:
R. Crump, N. Killingstad, T. McCanna, S. Caruso, A. Laorr, K., Mork,
E. Steinbergs, E. Wigley. With C. Mallory: N. Mahmud.  

We thank J. R. Stauffer for helpful comments on the manuscript.

This research has made use of NASA's Astrophysics Data System (ADS)
Abstract Service, and of the SIMBAD database, operated at CDS,
Strasbourg, France.  This research has made use of data products from
the Two Micron All-Sky Survey (2MASS), which is a joint project of the
University of Massachusetts and the Infrared Processing and Analysis
Center, funded by the National Aeronautics and Space Administration
and the National Science Foundation.  These data were served by the
NASA/IPAC Infrared Science Archive, which is operated by the Jet
Propulsion Laboratory, California Institute of Technology, under
contract with the National Aeronautics and Space Administration.  This
research has made use of the Digitized Sky Surveys, which were
produced at the Space Telescope Science Institute under U.S.
Government grant NAG W-2166. The images of these surveys are based on
photographic data obtained using the Oschin Schmidt Telescope on
Palomar Mountain and the UK Schmidt Telescope. The plates were
processed into the present compressed digital form with the permission
of these institutions. This research has made use of the NASA/IPAC
Extragalactic  Database (NED) which is operated by the Jet Propulsion 
Laboratory, California Institute of Technology, under  contract with
the National Aeronautics and Space Administration.

The research described in this paper was partially carried out at the
Jet Propulsion Laboratory, California Institute of Technology, under
contract with the National Aeronautics and Space Administration.

\appendix

\section{Comments on individual objects}

\subsection{073049.1-470209}
\label{app:073049.1-470209}

This object, 073049.1-470209, the westernmost in the set, is also the
reddest source in the set.  It is the YSO candidate farthest in the
upper right of  the IRAC color-color diagram (Fig.~\ref{fig:iracysos})
and is the  reddest and faintest YSO in [3.6] vs.\ [3.6]$-$[24] 
(Fig.~\ref{fig:3324ysos}). Based on its SED (Fig.~\ref{fig:seds1}),  it
is not surprising that there are no optical or even 2MASS 
counterparts; it very steeply falls from 24 \mum\ back to 3.6 \mum, 
and it seems that very deep integrations would be needed to  obtain
measurements of this object at shorter wavelength bands.  It is not
detected at 70 \mum, but it is in a high background region, so the 70
\mum\ limit is not very constraining. This type of SED can be found in
very young, very embedded, very low mass objects, but also in
extragalactic contaminants. Because of its proximity to the rest of
the Sa101 objects, it remains in the list, but we have given it the
lowest-quality grade (C).  Its steep SED means that it is one of three
Class I objects in our list.  Additional follow-up data are needed to
determine the true nature of this source.

\subsection{073049.8-465806}
\label{app:073049.8-465806}

This object, 073049.8-465806,  is relatively faint at 3.6 \mum, at
least in comparison  to the rest of the YSO candidates here. It
appears on the top edge of the clump of likely extragalactic sources
in the [3.6] vs.\ [3.6]$-$[24] plot (Fig.~\ref{fig:3324ysos}).  It is
only detected at $K_s$ in 2MASS, with limits at $J$ and $H$. Based on
the SED, it looks quite extinguished, e.g., there seems to be high
\av\ in the direction of this source.  It is located in close
(projected) proximity to other YSOs and YSO candidates; YSOs are also
frequently found near other YSOs.  It is not detected at 70 \mum, but
the limit is shallow enough that it does not provide a strong
constraint on the SED.  It has an SED characteristic of YSOs, and that
plus the apparently high \av\ plus its location in the cloud close to
other YSOs has yielded a ``B'' grade.  The steep rise of the SED from
4.5 to 24 \mum\ influences the classification slope fitting such that
it is one of three Class I objects in our list. Additional follow-up
data are needed.

\subsection{073053.6-465742}
\label{app:073053.6-465742}

Object 073053.6-465742 has 2MASS data, but no available optical data,
and just a limit at 70 \mum. The SED (Fig.~\ref{fig:seds1}) suggests
that there is high \av\ towards this source. Given our overly simple
modelling, there seems to be a small but significant excess at 8 \mum\
(better modelling is required to confirm this), and there is an
apparently large excess at 24 \mum.  Whenever an apparent excess is
seen only at 24 \mum, because the MIPS-24 camera has lower spatial
resolution than the shorter bands, there is a risk that the flux
density measured at 24 \mum\ is contaminated by source confusion, most
likely with a nearby background source, but it could also be with a
low  mass companion to the young star candidate.  There is only a
limit at 70 \mum\ to help constrain the SED, though it is located in
close (projected) proximity to other YSOs and YSO candidates. In this
case, the fact that the excess appears to be significant at at least 2
bands, its proximity to other nearby Sa101 sources, and the likelihood
that there is high \av\ towards this source suggests that it may be a
legitimate YSO, so we have placed it in the highest quality source bin
(grade ``A''), and it is a Class II.  Spectroscopy of this target is
required to determine whether or not it is a YSO.

\subsection{073057.5-465611=CG-Ha2}
\label{app:073057.5-465611}

CG-H$\alpha$2 was identified in Reipurth \& Pettersson (1993) as a
YSO.  It has counterparts at all bands we considered here, including
70 \mum.  In the optical  (Fig.~\ref{fig:vviysos}), it is in the locus
of young stars  above the 30 Myr isochrone, and it appears there to be
the lowest-mass object of the YSO candidates with optical data. The
spectral type given in Reipurth \& Pettersson (1993) is M2:, so that
is the model we have used in Figure~\ref{fig:seds1}. It also is one of
the stars with a clear $B$ excess in Fig.~\ref{fig:uvysos}, which can
also be noted in the SED itself. Such an excess is a characteristic of
active accretion.  Because it is a previously identified YSO, it
appears as grade ``A+'' in Table~\ref{tab:ourysos2}; it is a Class II.

\subsection{073106.5-465454}
\label{app:073106.5-465454}

073106.5-465454 is the northernmost YSO candidate within the Sa101
association. It has a 2MASS counterpart, but the SED
(Fig.~\ref{fig:seds1}) seems somewhat ``disjoint'' between the 2MASS
and IRAC portions. This could be from very high extinction, or
intrinsic stellar variability between the epochs of observation at the
NIR and MIR, or source confusion.  It is detected at 24 \mum\ but
there is only a limit at 70 \mum. We categorize it as a ``C''-grade
YSO candidate, with a ``flat'' SED class.  Additional data are needed
to confirm or refute the YSO status of this object.

\subsection{073108.4-470130}
\label{app:073108.4-470130}

This object, 073108.4-470130,  has an SED (Fig.~\ref{fig:seds1}) quite
consistent  with other known YSOs, although there are no optical data
and just a limit at 70 \mum.  If it is really a young star, it has a
significant excess in at least 8 and 24 \mum, and probably 5.8 \mum\
as well. We categorize it as a very high quality YSO candidate
(``A''), SED Class II. 

\subsection{073109.9-465750}
\label{app:073109.9-465750}

Like the prior object, 073109.9-465750 has an SED 
(Fig.~\ref{fig:seds1}) quite consistent  with other known YSOs, again
without optical data.  There is a limit at 70 \mum, and it is in a
high background region, so there is little constraint to the SED. 
Like the prior object (073108.4-470130), if it is really a young star,
it has a significant excess in at least 2 bands. We categorize it as a
very high quality YSO candidate (``A''), SED Class II. 

\subsection{073110.8-470032=CG-Ha3}
\label{app:073110.8-470032}

CG-H$\alpha$3 is another YSO identified in Reipurth \& Pettersson 
(1993), and has some photometry reported there as well. It is detected
in all bands considered here, $B$ through 70 \mum.  In the optical
(Fig.~\ref{fig:vviysos}), it is well within  the clump of YSO
candidates above the 30 Myr isochrone.  As seen in its SED
(Fig.~\ref{fig:seds1}), the optical and NIR data from Reipurth \&
Pettersson (1993) are quite consistent with the optical and NIR data
we report here, with some weak evidence for variability in the
optical.  Reipurth \& Pettersson (1993) report a type of K7, and that
is the model we have used in Fig.~\ref{fig:seds1}.  Like
CG-H$\alpha$2, the SED suggests that there may be some $B$-band excess
(most likely due to accretion); in Fig.~\ref{fig:uvysos}, it appears
as one of the objects apparently pushed above the main sequence due to
reddening.  Because it is a previously identified YSO, it appears as
grade ``A+'' in Table~\ref{tab:ourysos2}; it is a Class II.

\subsection{073114.6-465842}
\label{app:073114.6-465842}

This object can be seen at all four IRAC bands and both of the  MIPS
bands considered here.  It also has a 2MASS counterpart, but no
optical data. The SED (Fig.~\ref{fig:seds1}) suggests that there is
relatively high \av\ towards this object. Of the new YSO candidates
that appear to have some photometric points on the photosphere, this
object appears to have the highest  \av. If this object is actually a
young star, it has an IR excess in four bands. We categorize this as
another highest-quality (grade ``A'') YSO  candidate, with SED Class
``flat.'' Additional data are needed, including optical photometry.

\subsection{073114.9-470055}
\label{app:073114.9-470055}

Another new YSO candidate, 073114.9-470055 has a 2MASS counterpart but
no optical data and just a limit at 70 \mum\ (Fig.~\ref{fig:seds2}).
If it is a legitimate young star, it has an IR excess in more than two
bands. We categorize it as a grade ``A'' YSO candidate, SED Class II. 
Additional data are needed, including optical photometry.

\subsection{073121.8-465745=CG-Ha4}
\label{app:073121.8-465745}

CG-H$\alpha$4 is another YSO from  Reipurth \& Pettersson (1993), and
has some photometry reported there as well as  a spectral type of
K7--M0.  As discussed above, this object has a high-quality $J$
measurement in the 2MASS point source catalog, but the $HK_s$
measurements were flagged as having low-quality photometry.  We
accepted the low-quality measurements with a large uncertainty and
added them to the SED in Fig.~\ref{fig:seds2}. Given the the
normalization as seen  there, there could be a disk excess beginning
as early  as 3.6 \mum, but with a better measurement at the near-IR
and thus a better constraint on the location of the  photosphere, the
disk excess could start at a longer wavelength, and we suspect that it
probably does. The optical data we report, as compared with the
optical data from Reipurth \& Pettersson (1993), suggest substantial
intrinsic source variability, a common characteristic of young stars. 
In the optical color-magnitude diagram in Figure~\ref{fig:vviysos}, it
is apparently the oldest YSO of the set of YSO candidates above the 30
Myr isochrone.  This object's apparent intrinsic variability could
move it around in this diagram; moreover, the uncertain distance to
the CG4+Sa101 cloud moves this object above or below the 30 Myr
isochrone (as discussed above). This is the only previously-known YSO
in our set that is not detected at 70 \mum.  It appears in
Fig.~\ref{fig:uvysos} as having one of the smallest $B$-band excesses,
but the photometry as seen in Fig.~\ref{fig:seds2} suggests that
perhaps its placement in Fig.~\ref{fig:uvysos} could be improved.
Because it is a previously identified YSO, it appears as grade ``A+''
in Table~\ref{tab:ourysos2}; it is a Class II.

\subsection{073136.6-470013=CG-Ha5}
\label{app:073136.6-470013}

CG-H$\alpha$5 is reported as a K2--K5 in Reipurth \& Pettersson (1993).
It is detected at all available bands we discuss here.  It appears in
Figure~\ref{fig:uvysos} as having a $B$-band excess and subject to
high \av.    Because it is a previously identified YSO, it appears as
grade ``A+'' in Table~\ref{tab:ourysos2}; it is a Class II.

\subsection{073137.4-470021=CG-Ha6}
\label{app:073137.4-470021}

CG-H$\alpha$6, located very close to CG-H$\alpha$5, is reported as a
K7 in Reipurth \& Pettersson (1993). It is detected at all available
bands we discuss here, $B$ through 70 \mum. The infrared excess
appears to start around 8 \mum, suggesting a possible inner disk
hole; more detailed modelling is needed to be sure. It appears in
Figure~\ref{fig:uvysos} as having a $B$-band excess and subject to
high \av.   It is another ``A+'', Class II, in
Table~\ref{tab:ourysos2}.

\subsection{073143.8-465818}
\label{app:073143.8-465818}

This object, 073143.8-465818, is detected at optical through 24 \mum.
It is a high-quality YSO candidate (grade ``A''), and SED class
``flat.'' In the optical (Fig.~\ref{fig:vviysos}), it is the lowest
mass YSO candidate without a prior identification. It has a clear $B$
excess in Fig.~\ref{fig:uvysos}.  It is not detected at 70 \mum. 
Spectroscopy of this object is required to confirm/refute its YSO
status and obtain an initial guess at its mass.

\subsection{073144.1-470008}
\label{app:073144.1-470008}

073144.1-470008 is another high-quality YSO candidate (grade ``A'',
Class II) , detected in the optical through 24 \mum.  It too has a
clear $B$  excess in Fig.~\ref{fig:uvysos}. It is undetected at 70
\mum.

\subsection{073145.6-465917}
\label{app:073145.6-465917}


073145.6-465917 is the last of the YSO candidates in this list
associated  with Sa101.  It is a high-quality (grade ``A'') YSO
candidate, detected at optical through 24 \mum, but with only a limit
at 70 \mum. Based on the approximate SED fit in Fig.~\ref{fig:seds2},
the disk excess starts at 8 \mum, though additional  modelling (and a
spectral type) are required to be sure. In Fig.~\ref{fig:uvysos}, it
appears as having a $B$-band excess and subject to \av. It has an SED Class
of II.

%

\subsection{073326.8-464842=CG-Ha7}
\label{app:073326.8-464842}

CG-H$\alpha$7 is the only previously known YSO in the CG4 region
covered by our maps.  Reipurth \& Pettersson (1993) report a K5
spectral type; their optical data are quite consistent with the
optical data we report. We detect this object at all available bands
reported here, $B$ through 70 \mum. In Fig.~\ref{fig:vviysos}, it is
the highest apparent mass YSO candidate, though reddening can strongly
influence this placement.  The disk excess  (Fig.~\ref{fig:seds2})
begins at the longer IRAC bands, suggesting a possible inner disk
hole.  It does not appear to have a $B$-band excess in
Fig.~\ref{fig:uvysos}. It is located far from any nebulosity
(Fig.~\ref{fig:i4withysos}), as noted by Reipurth \& Pettersson
(1993). It is an SED Class II.

\subsection{073337.0-465455}
\label{app:073337.0-465455}

Object 073337.0-465455 has counterparts at $J$ through 24 \mum; there
are no optical data available.  Its placement in the [3.6] vs.\
[3.6]$-$[24] diagram (Fig.~\ref{fig:3324ysos}) suggests that it is at
the edge of our detection limits; it appears at [3.6]$-$[24]$\sim$3.4,
[3.6]$\sim$14.  It is undetected at 70 \mum. If it is a legitimate
YSO, it has an excess in at least 3 bands. It is located just off the
northern edge of the globule. We place it in the ``Grade B'' bin, and
it is SED Class II.  Additional follow-up data are needed.

\subsection{073337.6-464246}
\label{app:073337.6-464246}

This object, 073337.6-464246, is detected at optical through 24 \mum,
but not at 70 \mum. It is the candidate seen in Fig.~\ref{fig:vviysos}
as very low in the optical color-magnitude diagram, and closest to the
IRAC photospheric locus of points in Fig.~\ref{fig:iracysos}. There is
a  marginal excess seen at 5.8 and 8 \mum, and then a larger excess at
24 \mum; as discussed above, excesses seen only (or primarily) in a
single point at 24 \mum\ can be a result of source confusion with
adjacent sources.  This source is also located very far from any
nebulosity, near the edges of some of our IRAC maps
(Fig.~\ref{fig:i4withysos}). We place this marginal candidate as a
grade ``C''; it is an SED Class II.  Additional data are needed.

%
%
%

\subsection{073406.9-465805}
\label{app:073406.9-465805}

073406.9-465805 is the only YSO candidate of our set located projected
onto the globule/elephant trunk.  It appears to also be projected onto
a bright rim at 8 \mum\ (Fig~\ref{fig:i4withysos}), so it is
potentially being revealed now by the action of the ionization front. 
There are no optical or 70 \mum\ counterparts, but there are
counterparts at $J$ through 24 \mum.  If it is a young star, there is
an IR excess at more than 2 bands. We place it as a Grade ``A,'' Class
II object, and additional data are needed.

\subsection{073425.3-465409}
\label{app:073425.3-465409}
\label{sec:extsrc}

073425.3-465409 is a very interesting source.  In the [3.6] vs.\
[3.6]$-$[24] diagram (Fig.~\ref{fig:3324ysos}), it is the brightest,
reddest object ([3.6]$\sim$11.7, [3.6]$-$[24]$\sim$8.2).  And, it is
located right on the ``lip'' of the globule
(Fig.~\ref{fig:i4withysos}), in a region where YSOs might be expected
to form.  None of the other ``fingers'' of molecular cloud have
apparent associations with infrared objects.  After the automatic
bandmerging described in \S\ref{sec:obs} above, this object did not
appear to have a 2MASS counterpart. However, in looking at the overall
shape and brightness of the SED, we suspected that it should have a
match in 2MASS.  Examining the images by hand, there is clearly a
source at this location visible in 2MASS and the POSS images. At the
POSS bands, it is distinctly fuzzy. It is also slightly resolved at
$J$; it has been identified with 2MASX J07342550-4654106. The position
given for this source in the 2MASS point source catalog is
2.1$\arcsec$ away from the position in the IRAC catalog, and the
position from the 2MASS extended source catalog is 1.98$\arcsec$ away
from the IRAC position, both of which are very large compared to most
of the rest of the catalog. If it is also slightly resolved at IRAC,
this could affect the positional uncertainty.  Certainly, the
structure of the molecular cloud around this source is complex and
could also have affected the position in the catalog. In any case,
manual inspection of the images ensures that the object is really the
same in the two catalogs, so the $JHK_s$ photometry from the extended
source catalog was attached to this Spitzer source.  The SED
(Fig.~\ref{fig:seds3}) rises steeply at the long wavelengths, but it
is unfortunately just off the edge of the 70 \mum\ maps. 

Given this object's SED and location, we have classified this as a
high-quality (grade ``A'') YSO candidate, with SED Class I. Additional
data are required to confirm or refute this object's status.

\subsection{073439.9-465548}
\label{app:073439.9-465548}

073439.9-465548, the last object in our list, has counterparts at $J$
through only 8 \mum, and as such, it has a sparse SED. (It is off the
edge of both the 24 and 70 \mum\ maps.) It is located just to the East
of the end of the globule, consistent with it having been relatively
recently uncovered.  The shape of the IRAC portion of the SED is a
little different than that for other objects in this set; despite the
negative overall slope of the IRAC points, none of the IRAC points
seem to be photospheric (compare to other SEDs in
Fig.~\ref{fig:seds1}--\ref{fig:seds3}), and there is even a very
slight negative curvature (the slope between 3.6 and 4.5 \mum\ is
slightly shallower than that between 4.5 and 5.8 \mum).  On the basis
of experience looking at many hundreds of SEDs for YSOs and contaminants
(Rebull \etal\ 2010, 2011), we have some reservations about the shape
of this SED; without additional photometric data, it is hard to give
this object a high grade. We give this one a ``C'' grade; it is a
Class II.

\end{document}